\documentclass[sigconf,nonacm]{acmart}

\usepackage{enumitem}

\usepackage{caption}
\newlength\myindent
\newcommand\bindent{%
  \begingroup
  \setlength{\itemindent}{\myindent}
  \addtolength{\algorithmicindent}{\myindent}}
\newcommand\eindent{\endgroup}

\newlength\wqindent
\newcommand\bwqindent{%
  \begingroup
  \setlength{\itemindent}{\wqindent}
  \addtolength{\algorithmicindent}{\wqindent}}
\newcommand\ewqindent{\endgroup}

\newlength\cindent
\AtBeginDocument{%
  \captionsetup[figure]{labelfont=normalfont,textfont=normalfont}
}

\AtBeginDocument{%
  }

\usepackage[linesnumbered,ruled,vlined]{algorithm2e}
\usepackage{pythonhighlight}
\usepackage{xcolor}
\usepackage{tabularx}
\usepackage{booktabs}
\usepackage{subcaption}
\usepackage{pifont}
\usepackage{hyperref}

\usepackage{algorithm}
\usepackage{algorithmic}

\newcommand{\fram}{\textsc{\textit{Flying Serving}}\xspace}
\definecolor{wq}{rgb}{0.6, 0.4, 0.8}
\definecolor{sw}{rgb}{0.4, 0.57, 0.88}
\begin{document}


\title{FLYING SERVING: On-the-Fly Parallelism Switching for Large Language Model Serving}

\makeatletter
\let\ACM@orig@affiliation\affiliation
\renewcommand{\affiliation}[1]{%
  \ACM@orig@affiliation{#1\country{}}%
}
\makeatother

\author{Shouwei Gao}
\affiliation{%
  \institution{Oregon State University}
}
\email{gaosho@oregonstate.edu}

\author{Junqi Yin}
\affiliation{%
  \institution{Oak Ridge National Laboratory}
}
\email{yinj@ornl.gov}

\author{Feiyi Wang}
\affiliation{%
  \institution{Oak Ridge National Laboratory}
}
\email{fwang2@ornl.gov}

\author{Wenqian Dong}
\affiliation{%
  \institution{Oregon State University}
}
\email{wenqian.dong@oregonstate.edu}

\renewcommand{\shortauthors}{Gao et al.}


\begin{abstract}
Production LLM serving must simultaneously deliver high throughput, low latency, and sufficient context capacity under non-stationary traffic and mixed request requirements. Data parallelism (DP) maximizes throughput by running independent replicas, while tensor parallelism (TP) reduces per-request latency and pools memory for long-context inference. However, existing serving stacks typically commit to a static parallelism configuration at deployment; adapting to bursts, priorities, or long-context requests is often disruptive and slow.
We present \fram, a vLLM-based system that enables online DP-TP switching without restarting engine workers. \fram makes reconfiguration practical by virtualizing the state that would otherwise force data movement: (i) a zero-copy \textit{Model Weights Manager} that exposes TP shard views on demand, (ii) a \textit{KV Cache Adaptor} that preserves request KV state across DP/TP layouts, (iii) an eagerly initialized \textit{Communicator Pool} to amortize collective setup, and (iv) a deadlock-free scheduler that coordinates safe transitions under execution skew. Across three popular LLMs and realistic serving scenarios, \fram improves performance by up to $4.79\times$ under high load and $3.47\times$ under low load while supporting latency- and memory-driven requests.
\end{abstract}

\maketitle

\section{Introduction}

Large language models (LLMs) have become a primary workload in modern AI services, powering interactive assistants, code generation, and reasoning-heavy applications~\cite{team2023gemini,team2025kimi,brown2020language,li2022competition,touvron2023llama, stojkovic2024dynamollm}. At production scale, \emph{inference} often dominates both user experience and operating cost: providers must deliver responsive, streaming outputs under strict latency expectations while keeping expensive accelerator fleets highly utilized~\cite{rajbhandari2025arctic,pope2023efficiently, lai2025tokenscale}. This is not a hypothetical concern. For example, Perplexity reports serving over 435M search queries per month, where each user query triggers multiple inference requests in the backend~\cite{nvidia2024perplexity}. Large providers report similarly Internet-scale LLM demand: SageServe characterizes Microsoft Office 365 workloads with over 10M LLM requests per day across regions and time~\cite{jaiswal2025sageserve}, while BurstGPT releases production traces from regional Azure OpenAI GPT services capturing burstiness and heterogeneity over long periods~\cite{wang2024burstgpt}. Cloud provider guidance similarly emphasizes that inefficient inference configurations can lead to ``skyrocketing costs,'' motivating careful performance engineering and benchmarking~\cite{azure2024opt_infer,dynamollm2024}.

This operational reality is increasingly captured by benchmarking practice. MLPerf Inference evaluates \emph{latency-bounded throughput}: reported throughput is only meaningful if the system satisfies explicit latency constraints under online arrivals~\cite{liu2025greenllm, reddi2019mlperf,li2025adaserve, mlperf_inference_docs}. For example, MLPerf specifies constraints in terms of time-to-first-token (TTFT) and time-per-output-token (TPOT), and recent releases explicitly introduce interactive, low-latency LLM benchmarks to reflect user-facing responsiveness requirements~\cite{wang2024burstgpt, nguyen2025llm_benchmarking, mlperf_inference_docs,mlcommons2025llm_v5}. More broadly, provider-facing benchmarking guidance frames deployment cost as a function of how many queries the system can serve per second \emph{while remaining responsive}~\cite{nvidia2025llm_benchmarking}. Together, these trends highlight a central systems objective for LLM serving: maximize throughput subject to end-to-end latency constraints, rather than optimizing either metric in isolation.

Current distributed inference stacks, however, force architects into a static choice between parallelism strategies. TP shards model weights across GPUs to reduce per-request latency and accommodate large models or long contexts under tight SLOs~\cite{shoeybi2019megatron}. DP replicates the model to maximize aggregate throughput by serving independent requests concurrently. The problem is rigidity: real serving workloads are non-stationary, with bursty arrivals, heterogeneous request lengths, and mixed priority classes, so a single fixed strategy is inevitably suboptimal~\cite{hidayetoglu2025shift}. TP-centric deployments can under-provision concurrency during bursts and suffer queue buildup, while DP-centric deployments can inflate tail latency and struggle to sustain long-context sessions due to memory pressure from weights and KV cache. While other parallelization strategies, such as Pipeline Parallelism (PP)~\cite{huang2019gpipe} and Expert Parallelism (EP)~\cite{liu2024deepseek}, are utilized in specific scenarios, particularly for Mixture-of-Experts (MoE) models~\cite{gptoss_modelcard, liu2024deepseek}, DP and TP serve as the fundamental building blocks of distributed inference. Crucially, TP and DP are universally applicable and orthogonal to these specialized methods, allowing them to be compatible with other strategies as needed.

An ideal serving system should \emph{adapt} its parallelism online. During low load, it coalesces devices into TP groups to minimize tail latency. During bursts, it should expand into DP replicas to increase concurrency and drain queues. Realizing this policy is technically challenging because DP and TP differ in (i) how they place and access weights, (ii) how they allocate and index KV cache, and (iii) how they form communication groups. Intuitive switching between modes requires weight replication or reload, KV cache migration or duplication, and disruptive reconfiguration of communication state, which costs that erase the benefit of adapting under load.

To address these challenges, we present \fram, a distributed LLM serving system that dynamically switches between DP and TP \emph{without} model replication, KV cache transfer, or prohibitive reconfiguration pauses. \fram supports both execution modes from a unified in-memory organization: it (1) enables zero-redundancy weight management across DP/TP, (2) maintains a shared KV cache pool that remains stable across modes, and (3) activates the appropriate communication topology on demand. A workload-aware scheduler coordinates switching under execution skew and mixed priorities.


Specifically, we make the following contributions in this paper:
\begin{itemize}
\item We propose \fram, a vLLM-based serving system that virtualizes each engine as a DP replica and uses a global task pool plus a workload-aware scheduler to dynamically \emph{merge} replicas into TP groups (and \emph{split} them back), enabling adaptation to bursty demand, mixed priorities, and long-context requests.

\item We introduce a switching substrate that makes DP$\leftrightarrow$TP transitions lightweight: a \textit{Model Weights Manager} that loads weights once and realizes TP via logical shard activation (no tensor movement), a \textit{KV Cache Adaptor} that maintains a single physical KV pool with constant-time remapping across modes, and a \textit{Communicator Pool} that activates the required control- and data-plane collectives on demand.

\item We implement \fram as a set of patches to vLLM~\cite{kwon2023efficient}, preserving core optimizations including PagedAttention~\cite{kwon2023efficient}, chunked prefill~\cite{agrawal2023sarathi}, and continuous batching~\cite{seesaw2025}.

\item We evaluated \fram under bursty, production-like workloads and achieved a speedup of 4.79x and 3.47x for high and low workloads, respectively.

\end{itemize}

\section{Background and Motivation}

\subsection{Transformer Architecture}

\begin{figure}[h]
  \centering
  \includegraphics[width=\linewidth]{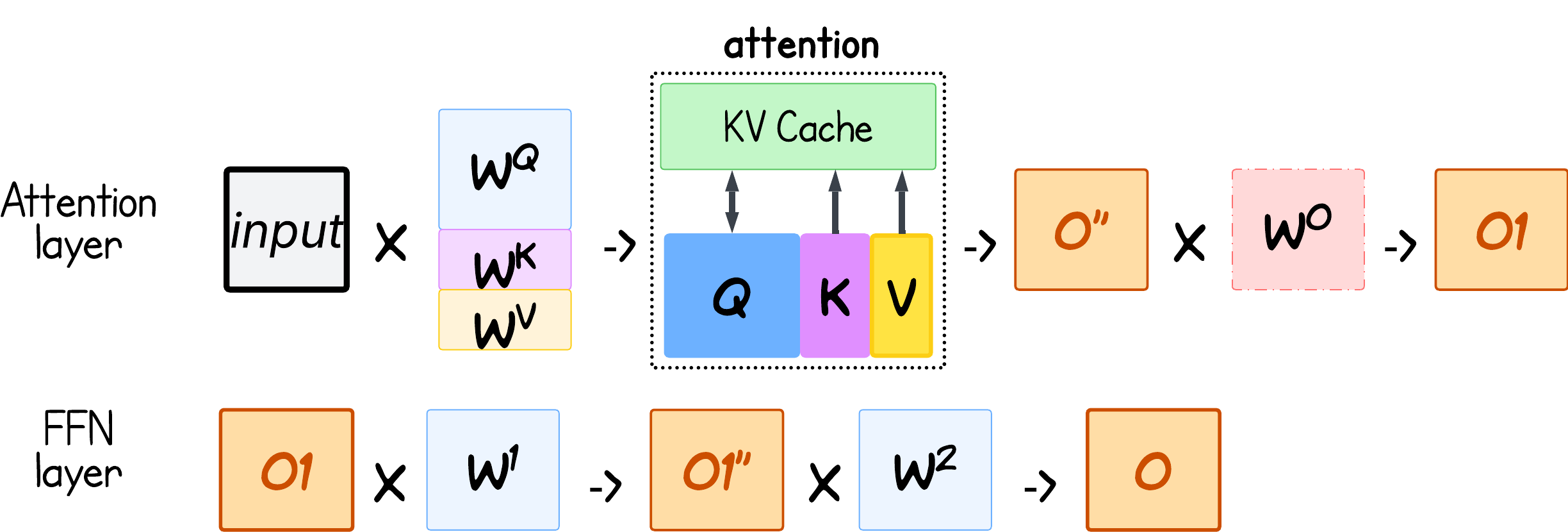}
  \caption{Attention/FFN operators (simplified).}
  \Description{xxx.}
  \label{fig:bg}
\end{figure}

Transformer inference is organized around a small set of high-impact linear operators whose shapes and execution modes govern both compute intensity and memory movement. As shown in Figure \ref{fig:bg}, each Transformer block consists of an Attention(MHSA or GQA~\cite{ainslie2023gqa}) operator followed by a Feed-Forward Network (FFN) operator. Although conceptually simple, these operators exhibit different tensor shapes and data dependencies across \textit{prefill}, \textit{batched decoding}, and \textit{single-token decoding}, which in turn drive the system-level behavior of LLM inference.

\textbf{Attention Layer.} The MHSA sublayer is built from four projection matrices, i.e., Query ($W_Q$), Key ($W_K$), Value ($W_V$), and Output ($W_O$), that apply GEMMs to the input hidden state. In efficient implementations, the $W_Q$, $W_K$, and $W_V$ projections are typically fused into a single GEMM operation applied to the input hidden state. 

When the model processes a sequence of tokens, the fused projection generates corresponding $q$, $k$, and $v$ vectors for each token in parallel. The resulting $k$ and $v$ vectors are appended to the KV cache, a persistent per-layer buffer. The attention mechanism then computes over the current and cached states to produce an intermediate output, which is finally projected by $W_O$ to produce $O^1$. 

\textbf{FFN Layer.} The FFN is composed of two large linear projections: an up-projection ($W^1$) and a down-projection ($W^2$), applied independently to each token. 
Despite their simplicity, these two GEMMs typically dominate FLOP count.



Crucially, this weight distribution dictates the topology of the KV cache. The KV cache consists of the intermediate key and value tensors generated by the $W_K$ and $W_V$ linear layers during the forward pass. Because these tensors are produced locally on the device holding the corresponding weight shard, the KV cache is inherently stored on the same device as the weights that generated it. Consequently, different parallelism configurations (e.g., varying the degree of TP) result in fundamentally different KV cache topologies and total capacities per device, creating a tight coupling between model weight distribution and memory management.





\begin{figure}[h]
  \centering
  \includegraphics[width=\linewidth]{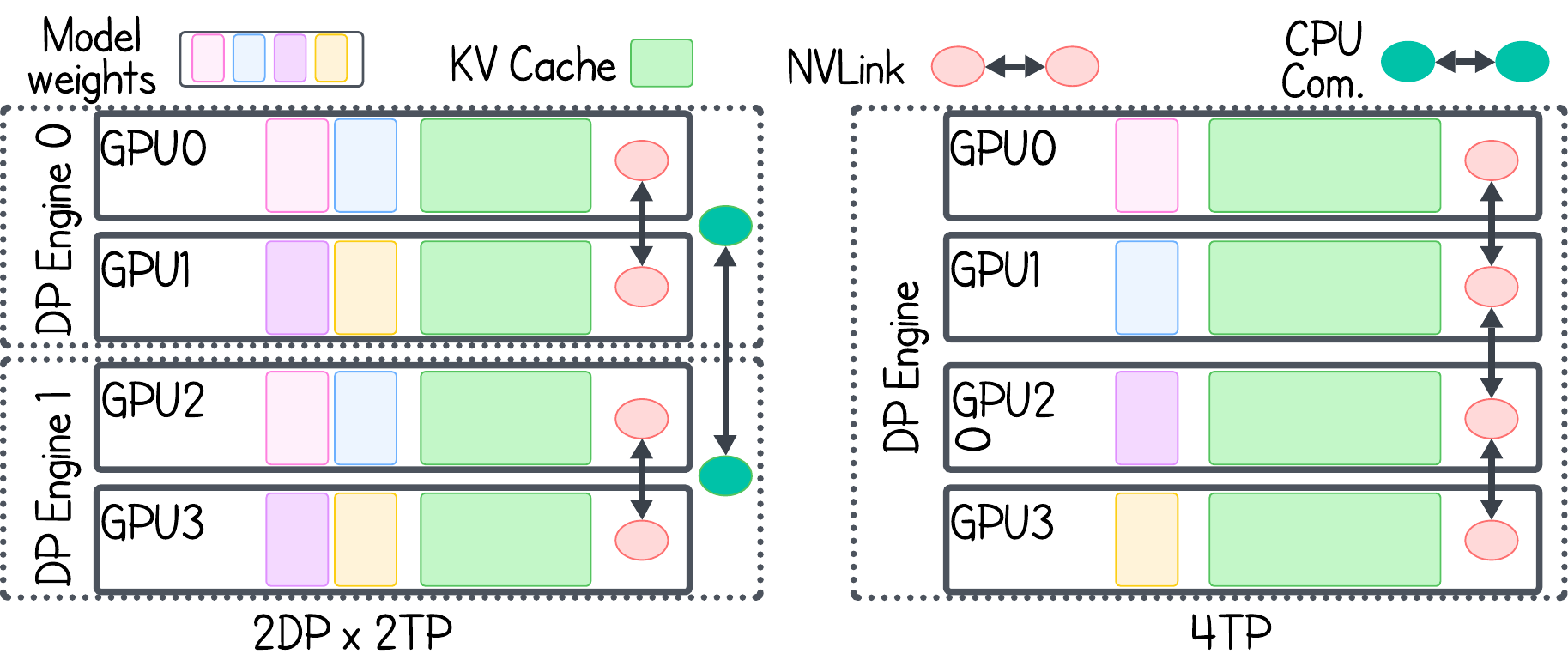}
  \caption{ Model layouts on a 4-GPU node in DP vs. TP.}
  \Description{TP vs DP.}
  \label{fig:dp_tp_comparison}
\end{figure}
\vspace{-10pt}
\subsection{Data Parallelism and Tensor Parallelism }

Serving large Transformer models requires distributing both computation and memory across multiple GPUs, and different workloads demand different forms of parallelism. DP and TP are two effective strategies for scaling large models and LLM serving. However, each strategy imposes its own requirements on weight layouts, KV cache organization, and communication patterns. These characteristics directly shape the design space explored in our framework design (Section~\ref{sec:system_design}): the feasibility of dynamic reconfiguration, the cost of state migration, and the limits on context length and throughput.

\begin{figure*}[ht]
  \centering
  \includegraphics[width=\textwidth]{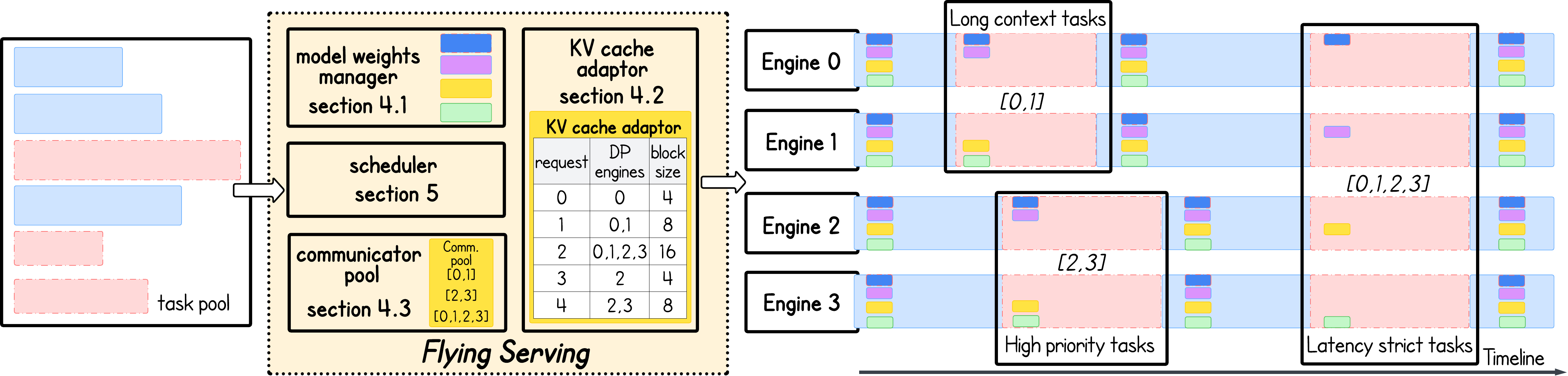}
  \caption{Overview of the \fram architecture. The system functions as a middleware layer that orchestrates multiple engine workers, enabling dynamic transitions between DP and TP. The timeline on the right illustrates how the system adapts to different request types, such as high-priority, long-context, and latency-strict tasks, by reconfiguring workers from independent DP instances into cooperative TP groups on the fly.}
  \Description{System architecture diagram}
  \label{fig:overall}
\end{figure*}

Figure~\ref{fig:dp_tp_comparison} contrasts DP and TP on a 4-GPU cluster, and Figure~\ref{fig:overall} illustrates how our system leverages both modes by smoothly transitioning between them.

\subsubsection{Data Parallelism: Independent Engines with Replicated State.} DP treats each group of GPUs as a standalone serving engine. In the $2\mathrm{DP} \times 2\mathrm{TP}$ example in Figure~\ref{fig:dp_tp_comparison}, the cluster forms two identical engines: (GPU0, GPU1) and (GPU2, GPU3). Each engine holds replicated weight shards, runs its own inference loop, and serves disjoint request sets.

(1) Memory duplication: Because each engine replicates the full model (or TP shard), the effective memory available for the KV cache shrinks. 

(2) Minimal communication: Engines operate independently and synchronize only at coarse granularity (e.g., request scheduling or step alignment as in vLLM~\cite{kwon2023efficient}), making DP suitable for low-overhead, high-throughput serving.

(3) Natural throughput scaling: Since each engine emits tokens independently, total tokens/sec increases linearly with the number of DP engines, even though individual requests are confined to the compute resources of one engine.

Together, DP excels for short-context or latency-sensitive traffic; however, it falls short for long-context or compute-heavy workloads that demand more memory or parallel compute than a single DP engine can provide.



\subsubsection{Tensor Parallelism: Distributed Operators for a Single Request.} TP partitions each linear projection operator in both the attention and FFN layers, across multiple GPUs. In the pure $4\mathrm{TP}$ configuration (Figure~\ref{fig:dp_tp_comparison}), all four GPUs jointly execute each operator invocation. This shifts the parallelization granularity from requests (DP) to operators (TP).

(1) Aggregated memory for weights and KV cache: Sharding weights across devices eliminates duplication. The KV cache inherits this sharding, effectively increasing the usable memory budget by (roughly) the TP degree. This offers longer context windows than DP.

(2) Lower latency for compute-bound workloads: Prefill and batched decoding benefit from parallel operator execution across GPUs, reducing per-request latency.

(3) High communication intensity: Every operator invocation requires synchronizing intermediate tensors among all TP workers. This mandates high-bandwidth interconnects (e.g., NVLink) and creates non-trivial \textit{reconfiguration challenges} if TP groups are changed dynamically.

TP is necessary for long-context or high-compute requests, but expensive to maintain during heterogeneous workloads.


\subsection{User Scenarios Motivating Dynamic DP-TP }
\label{sec:user_scenarios}

Existing systems typically commit to a fixed parallelism strategy: either primarily DP or primarily TP. However, real serving workloads are heterogeneous in time, priority, and context length, and no single static configuration can serve all of these efficiently. We highlight three common scenarios where dynamically switching between DP and TP yields substantial benefits.

\textbf{Use Case 1: Adapting to Time-Varying Load.}
Serving traffic is highly non-stationary. During low-load windows, the serving system backend has spare GPU capacity and can afford to over-provision resources per request. In this regime, forming TP groups is beneficial: the system aggregates GPUs to accelerate individual requests, lowering latency and improving the tail quality of service (QoS). When the workload spikes and request queues build up, the objective flips: the system should dissolve TP groups into more DP engines to maximize aggregate throughput and drain the queue. A dynamic DP-TP system can track load and continuously rebalance between "few fast TP engines" and "many DP engines" rather than committing to one extreme.

\textbf{Use Case 2: Priority-Aware Service Differentiation.}
Multi-tenant deployments often enforce tiered service levels, where a subset of requests (e.g., premium users or critical applications) carry stricter latency SLOs. This is analogous to the scheduler's Quality of Service (QoS) mechanism in HPC facilities, where certain jobs are granted higher scheduling priority. Under a static configuration, either all requests benefit from TP (wasting resources on low-priority traffic), or all are constrained to DP (violating high-priority SLOs). With dynamic DP-TP, the scheduler can selectively assign high-priority requests to TP groups to minimize their latency, while routing best-effort traffic to DP engines that emphasize throughput. This enables fine-grained, per-request allocation of parallelism rather than a global, one-size-fits-all policy.

\textbf{Use Case 3: Serving Long-Context Requests.}
The maximum context length a single serving engine can handle is bounded by its available GPU memory, dominated by the KV cache. In a pure DP configuration, if an engine supports sequences up to length $L$, a request with length $L + m$ will trigger an out-of-memory failure. Dynamically merging multiple DP engines into a TP group effectively pools their memory: the KV cache is sharded across all participating GPUs, increasing the usable capacity by approximately the TP degree. This allows the system to "scale up" on demand to serve long-context requests that exceed the limits of any single engine, while retaining the ability to "scale out" via DP for shorter, throughput-oriented workloads.

\section{An Overview of \fram}  

Figure~\ref{fig:overall} shows \fram as a middleware between a global \textit{Task Pool} and a set of engine workers: in the static mode, engines run as independent DP instances; when a request requires tighter latency or more effective memory capacity, \fram intelligently merges multiple engines into a TP group and later dissolves them back to DP.

\textit{Execution substrate: DP engines as the base unit.}
In \fram, a single LLM engine is the fundamental \textit{DP instance}. Each engine encapsulates one or more \textit{Workers} mapped to one (or a fixed small set of) physical devices. Each worker maintains a persistent memory pool partitioned into (i) static storage for \textit{Model Weights} and (ii) dynamic storage for the \textit{KV Cache}. Incoming requests are first aggregated in a global \textit{Task Pool}; by default, engines pull tasks and execute independently in DP mode to maximize throughput and absorb bursty arrivals.

\textit{One control abstraction: forming and releasing TP groups.}
The key capability of \fram is to \emph{bind} a subset of DP engines into a cooperative TP group for a specific request, then \emph{release} those engines back to DP. This bind/release operation is the only switching primitive exposed to the scheduler; all complexity is contained in the middleware components that make switching cheap and safe at runtime:
\begin{itemize}
    \item \textbf{Model Weights Manager (Section~4.1)} provides a rank-consistent \emph{logical view} of weights under different TP degrees without requiring per-switch weight reloads.
    \item \textbf{KV Cache Adaptor (Section~4.2)} preserves a consistent physical KV allocation while remapping request-level KV layout across DP and TP execution, so request state remains valid through switches.
    \item \textbf{Communicator Pool (Section~4.3)} supplies ready-to-use communication groups for candidate engine sets, enabling immediate collective execution when engines are bound.
    \item \textbf{Scheduler (Section~5)} selects \emph{when} to bind/release and \emph{which} engines participate, ensuring deadlock-free coordination under execution skew.
\end{itemize}

\textit{How \fram serves the three workload scenarios.}
The timeline in Figure~\ref{fig:overall} illustrates how the scheduler uses the same bind/release primitive to satisfy different request types.
\textbf{(Use Case 1: workload pattern adaptation)} When load is high, \fram keeps engines in DP to maximize concurrency and drain the queue; under light load, it opportunistically forms TP groups to reduce per-request latency.
\textbf{(Use Case 2: priority-based differentiation)} High-priority tasks can trigger an immediate TP binding to obtain more compute per request and tighter latency, while normal tasks continue to execute on remaining DP engines.
\textbf{(Use Case 3: long-context scaling)} Long-context tasks that pressure KV capacity are routed to wider TP groups so that KV is effectively sharded across engines; once the long-context request progresses past the memory-critical phase, engines are released back to DP to recover throughput. 


\begin{figure}[t]
  \centering
  \includegraphics[width=\linewidth]{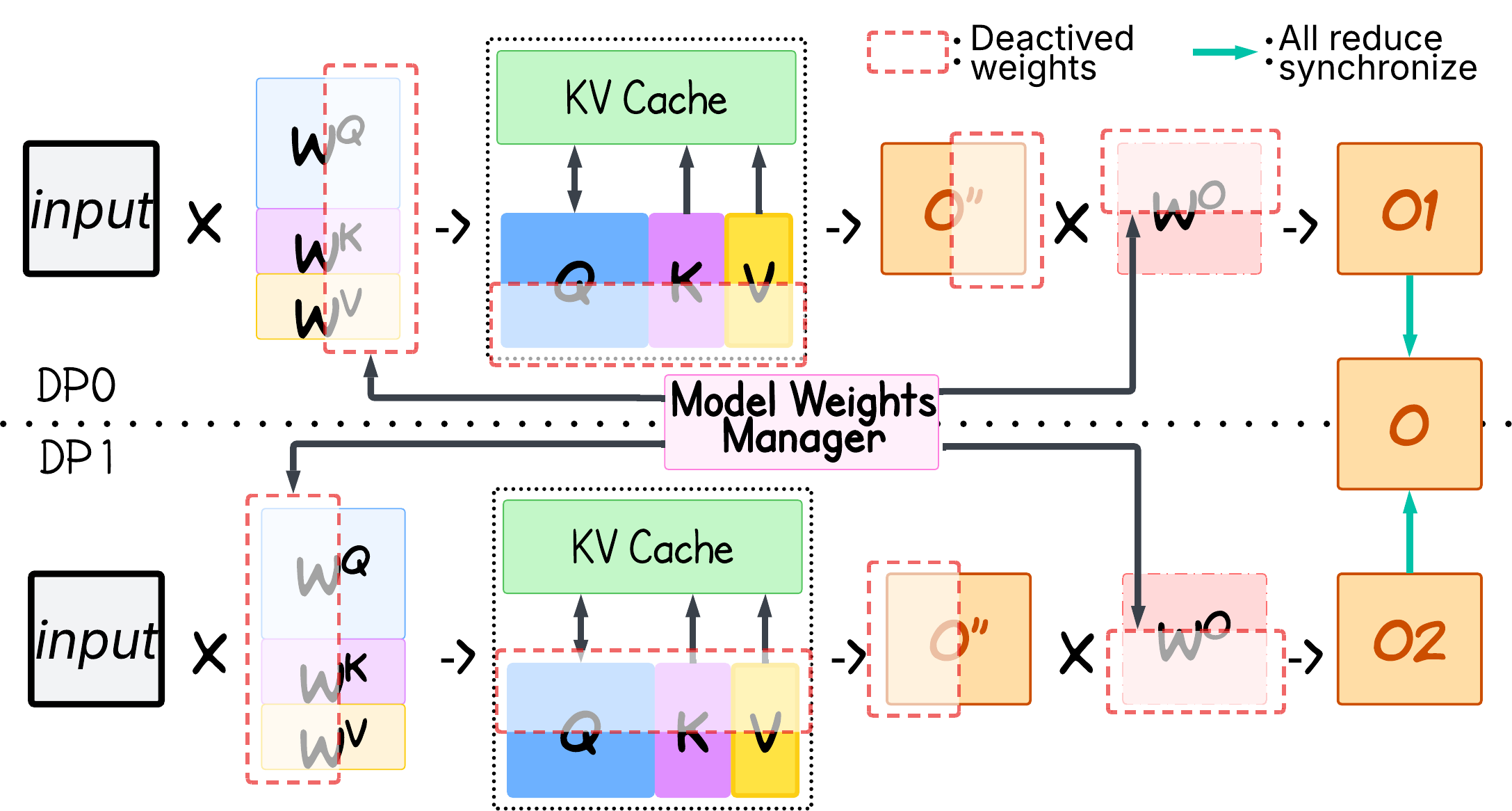}
  \caption{Model Weights Manager architecture for zero-copy DP/TP switching.}
  \Description{Diagram showing the Model Weights Manager.}
  \vspace{-1em}
  \label{fig:weights_manager}
\end{figure}

\section{Runtime Substrate for Dynamic Parallelism}
\label{sec:system_design}

\subsection{Model Weights Manager}
\label{sec:weights_manager}


Dynamic switching between DP and TP fundamentally changes how GPU workers consume model parameters. In DP, each engine requires a full replica of the model weights, whereas in TP, each engine accesses only a shard of each linear operator. Naively transitioning between these modes would require reallocating or reloading large parameter tensors, incurring transient memory and communication overhead. Such costs are incompatible with online serving and directly reduce the memory budget available for the KV cache.

The Model Weights Manager eliminates this reconfiguration bottleneck by decoupling \emph{logical} weight sharding from \emph{physical} weight storage. Its core invariant is that model parameters are loaded exactly once per engine and never physically moved thereafter. Mode switching is realized purely by changing which portions of an existing weight tensor are \emph{activated} for computation.

In Transformer architectures, the majority of parameters reside in linear layers, specifically the QKV ($W^Q, W^K, W^V$) projections and the Output projection ($W^O$) in Attention layers, as well as the Up, Gate, and Down projections in FFN (as referred in Figure~\ref{fig:bg}).

Figure~\ref{fig:weights_manager} illustrates our partitioning strategy using an Attention layer as an example. In a standard DP configuration, both Engine 0 and Engine 1 hold identical, full replicas of the weights ($W^Q, W^K, W^V$ and $W_O$). When the scheduler triggers a transition to a $2\times$ TP configuration, the system must logically partition these weights across devices to parallelize computation.

\subsubsection{Logical Weight Resharding}

Specifically, we adopt the Megatron-LM style parallelism~\cite{shoeybi2019megatron} to enabledynamic sharding:
\begin{itemize}[leftmargin=*]
    \item \textbf{Column-Parallel Layer ($W^{QKV}$):} We shard the fused $W^{QKV}$ projection along its output (column) dimension. In Figure~\ref{fig:weights_manager}, Engine~0 activates the first column slice while Engine~1 activates the second; all other columns in each local replica are \emph{deactivated} and excluded from computation. This partition maps cleanly to multi-head attention: each engine produces its local $Q$, $K$, and $V$ sub-tensors for a disjoint subset of heads, and thus requires no cross-engine communication for the $W^{QKV}$ projection.

    \item \textbf{Row-Parallel Layer ($W^{O}$):} We shard the output projection $W^{O}$ along its input (row) dimension. Each engine multiplies the full attention output with its local row slice, producing a partial projection ($O_1$ on Engine~0 and $O_2$ on Engine~1). A single \emph{all-reduce} then aggregates these partial results to form the final output activation $O$ for the layer.

\end{itemize}
This pattern ensures that only one synchronization step is required per pair of linear layers, maintaining high computational efficiency.

\subsubsection{Zero-Copy View Implementation}
A naive implementation of the above switching logic involves creating new shard tensors for each TP rank, which introduces duplicate data and causes memory allocation overhead. Instead, \fram implements a non-invasive patch to vLLM's \texttt{linear.py}\footnote{
vLLM implements its core linear operators (including tensor-parallel variants) in \texttt{vllm/model\_executor/layers/linear.py}~\cite{vllm-linear-doc}. Our patch extends this operator to accept a rank-aware tensor \emph{view} of the original weight matrix, allowing dynamic slicing along the sharding dimension without allocating new storage or modifying kernel code.} module to enable \textit{zero-copy}.

Specifically, when $m$ DP engines merge into an $m$-way TP group, the Manager assigns each engine a unique rank ID $r \in [0, m-1]$, and creates a \textit{view} of active weights on top of the corresponding full matrix:
\begin{equation}
    W_{active}^{(r)} = \text{View}(W_{full}, \text{dim}, r, m)
\end{equation}
where $\texttt{dim}$ defines the sharding dimension (columns for $W^{QKV}$, rows for $W^O$).

This approach guarantees that the active weights used for TP execution are contiguous in virtual memory but map to the existing physical memory of the DP replica. By avoiding physical data movement or duplication, \fram maximizes the available memory for the KV cache, distinguishing our approach from methods that rely on redundant storage or expensive reloading ~\cite{hidayetoglu2025shift, wu2023fastserve}.

\subsection{KV Cache Adaptor}
\label{sec:kv_adaptor}


The KV Cache Adaptor enables mode switching without migrating KV state or rebuilding the cache allocator. The core difficulty is that TP changes the per-device KV tensor shape: under DP, each worker stores KV tensors for the full hidden dimension, while under TP, each worker stores only a $1/p$ slice for TP degree $p$. In existing engines, these shapes are baked into the allocator at initialization; switching DP$\leftrightarrow$TP would therefore require reallocating the KV pool and repopulating metadata, which is too costly on the serving critical path.

\begin{figure}[ht]
  \centering
  \includegraphics[width=\linewidth]{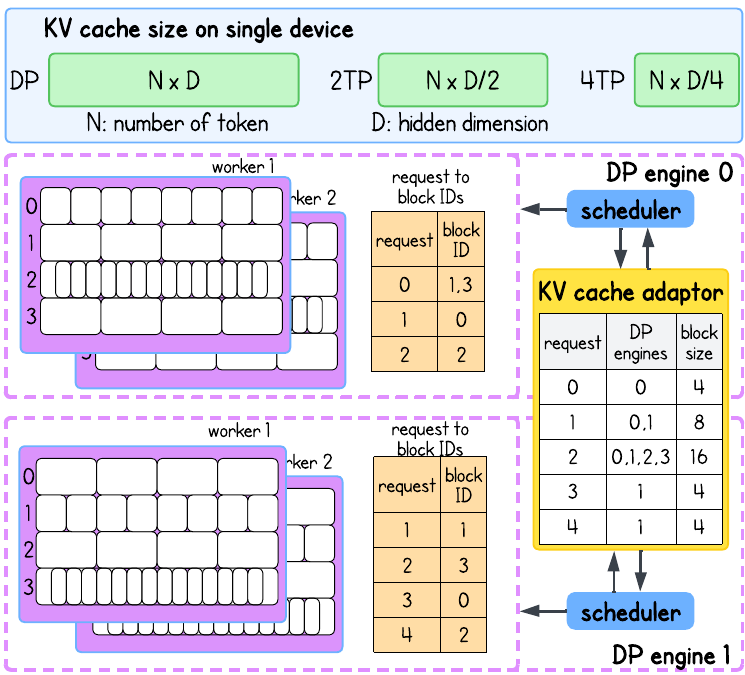}
  \caption{An example of KV cache adaptation in \fram. With per-token footprint shrinks with TP (DP: $N\!\times\!D$; 4TP: $N\!\times\!D/4$), to keep a fixed physical layout without reallocation, we scale block size inversely with TP: 4 tokens (DP), 8 (2TP), 16 (4TP), managed per request by the KV cache adaptor; physical memory of each block is unchanged. }

  \Description{KV Cache adaptation strategy.}
  \label{fig:kv_cache}
\end{figure}



\subsubsection{Challenge: Mode-Dependent KV Layouts}
DP and TP induce different per-device KV tensor shapes. Let $N$ be the number of cached tokens and $D$ the model hidden dimension. In DP mode, each device processes independent requests and must store the full hidden dimension for every token. The memory requirement per token on a single device is proportional to $D$. In TP mode, the model weights and intermediate tensors are sharded across $p$ devices. Consequently, the KV cache for a single token is partitioned, which means each device stores a slice of size $D/p$.

In existing serving engines, the KV allocator is initialized with a fixed block shape (determined by $D$ and a pre-set block size) derived from a single deployment configuration. 
Switching from DP to TP reduces the local tensor width from $D$ to $D/p$. If the system retains the original allocator settings, memory becomes misaligned and fragmented.
Reconfiguring the allocator to match the new shape requires rebuilding the memory pool. Such an operation is orders of magnitude slower than the per-step execution budget and would stall latency-critical requests. 

\subsubsection{Unified Memory via Adaptive Block Sizing}
To address this, \fram introduces the \textit{KV Cache Adaptor}, a middleware layer that dynamically manages the KV cache block. In vLLM's PagedAttention design, the KV cache is allocated as non-contiguous fixed-size physical blocks on the GPU and indexed via an OS-style logical mapping table. Instead of resizing physical memory blocks to fit the changing tensor shapes, we leverage the inverse relationship between the local hidden dimension $D$ and the token capacity per block $B$.


Concretely, the bytes consumed by a physical block are:
\begin{equation}
    M_{\text{block}} = B \cdot D_{\text{local}} \cdot P_{\text{size}},
\end{equation}
where $B$ is the number of tokens per block, $D_{\text{local}}$ is the per-device hidden dimension under the current mode, and $P_{\text{size}}$ is the element size. 
Our key insight is to keep $M_{\text{block}}$ constant across all modes to avoid reallocation.
We achieve this by inversely scaling the block size $B$ against the TP degree $p$.
Formally, for a given parallelism degree $p$, the local dimension becomes $D_{\text{local}}(p) = D/p$. To maintain memory alignment, we adjust the block size $B(p)$ such that:
\begin{equation}
\label{eqn:2}
B(p) = p \cdot B_{\text{base}}
\end{equation}
Where $B_{\text{base}}$ is the block size configured for the DP mode.

At runtime, as illustrated in Figure~\ref{fig:kv_cache}, the scheduler issues allocation and append requests in request space (e.g., request IDs), while the \textit{KV Cache Adaptor} resolves these requests to physical block IDs in the shared table. 
In DP mode, each DP engine maintains independent request streams; therefore, the adaptor can map a request to one or more block IDs that may be local to a specific engine (e.g., request $0 \mapsto \{1,3\}$, request $1 \mapsto \{0\}$, request $2 \mapsto \{2\}$ in the example table). When the system switches to TP, the adaptor preserves the physical residency of existing blocks and only updates the logical interpretation of each block (i.e., the effective block capacity $B(p)$) and the corresponding request-to-block mapping. In this way, mode transitions require neither KV-state migration nor allocator reinitialization: the adaptor performs constant-time metadata updates (capacity and indirection tables), while the underlying block pool and block IDs remain stable across DP/TP configurations.

\subsubsection{Implementation and Benefits}
We implemented this logic via a lightweight patch to the vLLM KV cache manager \footnote{ In vLLM, KV cache allocation and block-table management are implemented in the PagedAttention subsystem (e.g., \texttt{vllm/v1/core/kv\_cache\_manager.py} and the associated cache manager). Our patch extends this logic to allow per-request block sizing without modifying the underlying physical KV memory pool.}. The adaptor intercepts requests and dynamically updates a \textit{Logical Table} with the correct block size corresponding to the current parallelism mode of the worker. The worker informs both the KV cache kernel and the attention kernel of the stride and capacity for the specific request, ensuring correct memory indexing.

This design cleanly separates logical KV addressing from the physical block pool. A mode switch therefore reduces to metadata re-interpretation, i.e., updating per-request strides and block semantics, while leaving the allocator and underlying memory region unchanged. In effect, switching avoids KV migration and pool reinitialization, preserving cache residency and making reconfiguration lightweight. Moreover, when we consolidate multiple DP replicas into a TP group, each request can draw from the combined KV budget: the higher tokens-per-block density increases effective KV capacity, enabling longer contexts or larger batches without giving up the throughput benefits of the remaining DP replicas.

\subsection{Communicator Pool}
\label{sec:communicator}


Dynamic DP-TP switching requires changing which workers communicate and how. Existing serving stacks (including the vLLM infrastructure) typically construct a \emph{single, static} communication topology at startup, where CPU collectives (e.g., \textsc{Gloo}\textcolor{black}{~\cite{gloo}}) for control and GPU collectives (e.g., \textsc{NCCL}\textcolor{black}{~\cite{nccl}} over NVLink/PCIe) for tensor synchronization. Recreating or mutating these process groups at runtime is both slow (often tens of seconds) and brittle: in a multi-threaded inference loop, mismatched group membership can easily trigger collective hangs or deadlocks.

\begin{figure}[ht]
  \centering
  \vspace{-1em}
  \includegraphics[width=0.9\linewidth]{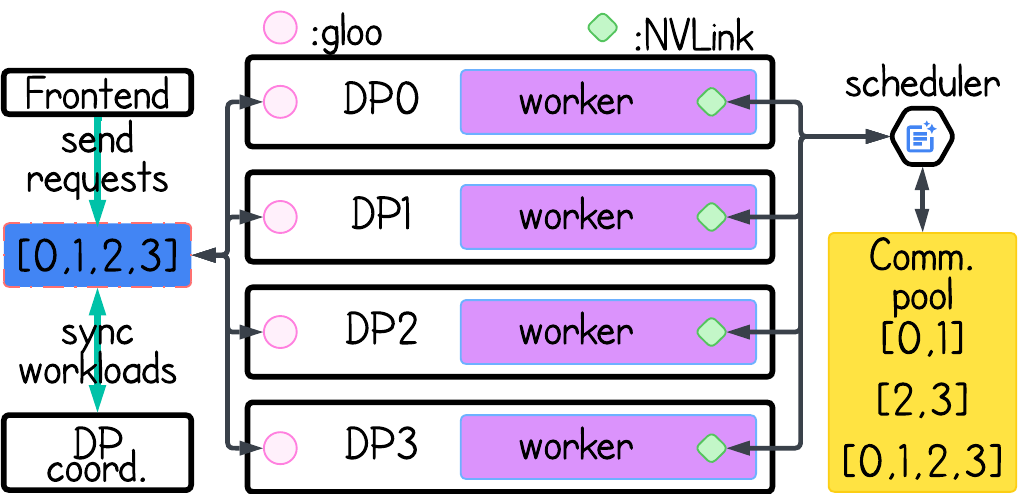}
\caption{\textbf{Two-plane communication in \fram.}
  \textbf{Left: Control plane (CPU--CPU).} The frontend routes requests and exchanges mode-switch signals with all DP engines via \textsc{Gloo}.
  \textbf{Right: Data plane (GPU--GPU).} The Communicator Pool pre-initializes all valid TP process groups (e.g., 2-way pairs and 4-way quartets) using \textsc{NCCL}. The scheduler activates the required group on demand, avoiding communicator creation on the request's critical path.}

  \Description{Communicator.}
  \label{fig:comm}
\end{figure}

To address this, \fram introduces a \textit{Communicator Pool} that decouples the initialization of communication channels from their execution. We abstract the communication architecture into two distinct planes: the Control Plane (CPU-CPU) and the Data Plane (GPU-GPU).


\subsubsection{Control Plane (CPU-CPU)}
The \textit{control plane} provides request distribution and global state synchronization. Under DP, each request is assigned to a single engine; under TP, a single request must be delivered to multiple engines that temporarily operate as one TP group. As shown in Figure~\ref{fig:comm} (left panel), we utilize the existing CPU-CPU communication pipes (implemented via \textsc{Gloo}) not only for distributing requests but also for synchronizing system state. 
In the default design, DP engines periodically synchronize their execution state via an all-reduce to detect globally unfinished requests\footnote{ \href{https://github.com/vllm-project/vllm/blob/29f7d9771569f26238d67cf6ea3a8792fb6a7792/vllm/v1/engine/core.py\#L1261}{\texttt{vllm/v1/engine/core.py/\#L1261}}}. We use a \textit{DP coordinator} piggybacks mode-switch signals (e.g., "\texttt{merge DP0 and DP1 into 2TP}") onto periodic synchronization heartbeats, ensuring that all participating engines observe the same transition point and apply it atomically.

\subsubsection{Data Plane (GPU-GPU) with Eager Initialization}
The dominant cost in reconfiguration is the GPU communicator setup (i.e., NCCL communication over NVLink or PCIe). Creating NCCL process groups and establishing connections can take seconds, which is unacceptable for mission-critical online serving. 

%
To enable dynamic switching without the latency of runtime initialization, \fram pre-allocates necessary communication groups at startup. However, naively enumerating all possible device combinations would lead to exponential resource overhead and memory exhaustion. Instead, we employ a \textit{topology-aware initialization} strategy that strictly targets physically contiguous device groups.
We build the pool in two steps:

\begin{itemize}
    \item[(1)] \textbf{ Topology-Aware Group Identification:} Given $N$ total DP engines (each using one GPU) and a set of supported TP degrees $\mathcal{P}$ (typically powers of two), we identify valid groups by partitioning the global rank space into contiguous segments. This constraint is critical because TP relies on high-bandwidth interconnects (e.g., NVLink), which typically connect adjacent GPU ranks.
    For example, with $N=4$ DP and $\mathcal{P}=\{2,4\}$, we do not generate strided or random combinations like $[0,2]$ or $[1,3]$. Instead, we restrict initialization to aligned, physically adjacent groups: $[0,1]$ and $[2,3]$ for 2TP, and $[0,1,2,3]$ for 4TP. This reduction ensures that the number of communicators scales linearly rather than exponentially.

    \item[(2)] \textbf{ Pre-initialization:} For every identified group, we invoke \texttt{torch.distributed.new\_group} (using the NCCL backend) during the system startup phase. The resulting communicator handles are cached in a hash map keyed by their member ranks (e.g., \texttt{Map<Tuple[int], Group>}).
    By pre-initializing only these topologically valid groups, \fram ensures that when the scheduler decides to merge DP workers into a TP instance (e.g., merging workers 0 and 1), the required communicator is already active and can be retrieved in $O(1)$ time, avoiding the costly overhead of runtime group creation.
\end{itemize}


At runtime, switching modes reduces to selecting the appropriate pre-built communicator handle from the pool and routing collectives through it. No new process groups are created on the critical path, which avoids both initialization latency and the deadlock risk of on-the-fly group reconstruction.
The memory overhead of keeping these pre-initialized communicators \emph{inactive} is small: in our measurements, each PyTorch distributed process group consumes $\sim$2\,MB of host memory.


\begin{figure*}[htb]
  \centering
  \includegraphics[width=\linewidth]{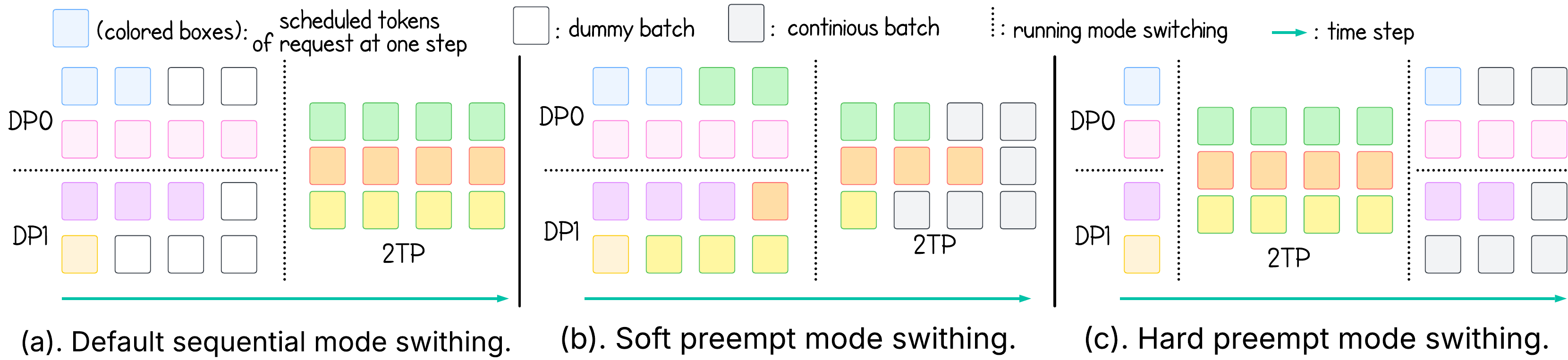}
  \caption{Comparison of Mode Switching Strategies in \fram. (a) \textbf{Default Sequential Switching:} All workers must idle until the longest-running DP request completes, leading to resource inefficiency. (b) \textbf{Soft Preempt:} Workers with available slots pre-execute TP requests in DP mode while waiting. Although this requires recomputing the KV cache upon switching to ensure correct layout, it saves significant decoding time. (c) \textbf{Hard Preempt:} High-priority TP tasks immediately interrupt active DP requests. DP tasks are resumed later without recomputation, leveraging the unified KV Cache Adaptor.}
  \Description{Timeline comparison of three switching modes.}
  \label{fig:switching_strategies}
\end{figure*}

\section{Dynamic Scheduler} 
\label{sec:scheduler}


The Dynamic Scheduler is the policy layer of \fram: it decides \emph{when} to form TP groups, \emph{which} engines participate, and \emph{how} to transition without stalling or deadlocking the serving loop. We implement the scheduler on top of vLLM v1\footnote{
vLLM v1 (SoTA version) was introduced as an \emph{alpha} release in January 2025 as a major re-architecture of vLLM's core engine, refactoring components such as the scheduler, KV cache manager, workers, sampler, and API server while reusing substantial parts of the v0 codebase (e.g., model implementations, GPU kernels, and the distributed control plane).}~\cite{vllm-v1}.
The scheduler's design is guided by two principles: (i) all engines participating in a TP step must observe an identical request order, and (ii) mode transitions must occur only at globally agreed safe points to avoid mismatched collectives.

\begin{algorithm}
\caption{Dynamic Scheduling \& Mode Switching Protocol}
\label{alg:scheduler}
\begin{algorithmic}[1]
\REQUIRE Base KV parameters: block size $B_{base}$; KV heads $H_{base}$;
\STATE Start with all engines in DP mode ($N_{eng}\leftarrow 1$);\\
\WHILE{True}
\bindent
\STATE \ding{182} Input Processing:\\
\STATE \hskip1em $Q_{in}\leftarrow$\texttt{ProcessInputSocket}();\\

\STATE \ding{183} Global Synchronization:\\
\STATE \hskip1em $Q_{wait}\leftarrow$\texttt{SyncWorkload}(\textit{peers});\\

\STATE \ding{184} Resource Scheduling:\\
\STATE \hskip1em $Q_{work}\leftarrow \emptyset$; 
\STATE \hskip1em $Flag_{SetTP}\leftarrow False$; 
\STATE \hskip1em $Flag_{ResetTP}\leftarrow False$; 
\STATE \hskip1em $N_{tp}\leftarrow 1$;\\
\FOR{$req \in Q_{wait}$}
    \bwqindent
    \STATE \ding{184} Mode Determination:\\

    \IF{$req.\textit{mode}=\text{TP}$}
        \bwqindent
        \STATE  $N_{eng}\leftarrow req.\textit{num\_engines}$; 
        \STATE $N_{tp}\leftarrow N_{eng}$; 
        \STATE $Flag_{SetTP}\leftarrow True$;\\
    \ewqindent
    \ELSE
     \bwqindent
        \STATE $N_{eng}\leftarrow 1$; 
         \STATE $Flag_{ResetTP}\leftarrow True$;\\
             \ewqindent
    \ENDIF

    \STATE \ding{185} KV Parameterization and Allocation: \\
    \STATE $B_{req}\leftarrow B_{base}\cdot N_{eng}$; 
       \STATE  $H_{req}\leftarrow H_{base}/N_{eng}$;\\
    \STATE $Blocks\leftarrow$\texttt{KVCacheMgr.Allocate}($req,B_{req},H_{req}$);\\
    \STATE  Append $(req.\textit{id},Blocks)$ to $Q_{work}$;\\
 \ewqindent
\ENDFOR

\STATE \ding{186} Mode Signaling (Collective RPC):\\
\IF{$Flag_{SetTP}=True$}
\bindent
    \STATE  \texttt{RpcBroadcast}
    (\texttt{"set\_TP\_mode"}, args=$N_{tp}$);\\
\eindent
\ELSIF{$Flag_{ResetTP}=True$}
\bindent
    \STATE \texttt{RpcBroadcast}(\texttt{"reset\_TP\_mode"});\\
\eindent
\ENDIF

\STATE \ding{187} Model Execution:\\
\STATE \texttt{RpcBroadcast}(\texttt{"execute\_model"}, args=$Q_{work}$);\\
\STATE  \texttt{PublishOutput}($Output$);\\
 \eindent
\ENDWHILE
\end{algorithmic}
\end{algorithm}

\subsection{The Base Execution Flow}

The scheduler runs as a centralized event loop that coordinates $K$ DP engines. Engines execute independently in DP mode by default; when a request is designated for TP, the scheduler temporarily coalesces a subset of engines into a TP group. Algorithm~\ref{alg:scheduler} summarizes one scheduling iteration, with six steps:


\textbf{Step \ding{182} -- Input Processing.}
At each iteration of the scheduling loop, the scheduler ingests newly arrived requests from the input socket and appends them to the local input queue $Q_{in}$ by invoking \texttt{ProcessInputSocket}(). This step performs no scheduling or mode decisions and serves solely to collect pending requests for subsequent coordination. 

\textbf{Step \ding{183} -- Global Synchronization.}
To ensure consistent task ordering and TP decisions across all engines, the scheduler synchronizes workloads with peers by invoking \texttt{SyncWorkload}(\textit{peers}), which returns a globally agreed waiting queue $Q_{wait}$. All subsequent scheduling and mode-selection logic operates exclusively on $Q_{wait}$, guaranteeing that engines participating in a TP step observe the same request sequence. 

\textbf{Step \ding{184} -- Resource Scheduling and Mode Determination.}
The scheduler initializes an empty worklist $Q_{work}$, clears the mode-switch flags $Flag_{SetTP}$ and $Flag_{ResetTP}$, and initializes the TP width variable $N_{tp}$. It then iterates over each request $req \in Q_{wait}$ to determine the execution mode. If $req.\textit{mode}=\text{TP}$, the scheduler sets $N_{eng}$, updates $N_{tp}$, and raises $Flag_{SetTP}$. Otherwise, it resets execution to DP mode by setting $N_{eng}=1$ and raising $Flag_{ResetTP}$. 

\textbf{Step \ding{185} -- KV Parameterization and Allocation.}
For each request, the scheduler derives mode-dependent KV cache parameters based on the selected engine width:
\begin{subequations}
\label{eqn:3}
\begin{align}
B_{req} = B_{base}\times N_{eng}, \\
H_{req} = H_{base}/N_{eng}.
\end{align}
\end{subequations}
Using these parameters, it invokes \texttt{KVCacheMgr.Allocate} to allocate KV cache blocks and appends the resulting tuple $(req.\textit{id},Blocks)$ to the worklist $Q_{work}$. 

\textbf{Step \ding{186} -- Mode Switching via Collective RPC.}
After processing all requests in the current iteration, the scheduler applies any required mode transition using collective RPCs. If $Flag_{SetTP}$ is true, it broadcasts \texttt{"set\_TP\_mode"} with argument $N_{tp}$; otherwise, if $Flag_{ResetTP}$ is true, it broadcasts \texttt{"reset\_TP\_mode"}. These broadcasts atomically configure the active communicator and execution mode across all engines before model execution. 

\textbf{Step \ding{187} -- Model Execution and Result Publication.}
With execution modes and KV cache allocations finalized, the scheduler invokes \texttt{RpcBroadcast}(\texttt{"execute\_model"}, args=$Q_{work}$) to execute one inference step collectively across engines. The generated outputs are then published via \texttt{PublishOutput}(), and the scheduler proceeds to the next iteration of the loop.

\subsection{Adaptive Mode Switching Strategies}

A central challenge in dynamic DP--TP execution arises from the service objectives of different user scenarios (Section~\ref{sec:user_scenarios}). Real-world serving workloads are non-stationary (Use Case~1), contain a mix of latency-critical and best-effort requests (Use Case~2), and occasionally demand transient scaling of memory capacity for long-context inference (Use Case~3). In all three cases, the scheduler must react to mode-switch opportunities under execution skew: different DP engines reach scheduling boundaries at different times due to heterogeneous request lengths.

\subsubsection{Default Sequential Switching.} Figure~\ref{fig:switching_strategies}(a) illustrates the baseline approach. When a TP request arrives, all DP engines wait for the longest-running request (the straggler) to complete before switching modes. While this strategy preserves correctness, it conflicts with the goals of Use Case~1 and Use Case~2: idle engines waste available capacity during load fluctuations, and latency-sensitive requests may be delayed by unrelated best-effort traffic. As a result, naive switching under-utilizes GPUs and introduces tail latency in practice.



\subsubsection{Soft Preempt (Throughput-Oriented).}
\textit{Soft preempt} addresses scenarios where throughput and load adaptability are the primary objectives (as in Use Case~1 and parts of Use Case~3). When a TP request is pending but some engines are still occupied with DP workloads, idle engines are allowed to speculatively execute the TP request in DP mode (Figure~\ref{fig:switching_strategies}(b)).  This leverages otherwise unused compute cycles instead of blocking on stragglers.

Although speculative execution produces a KV cache layout incompatible with TP, the trade-off is favorable. Decoding is typically memory-bound, while recomputation under TP is compute-bound and parallelized across multiple GPUs. By advancing generation during the waiting period, Soft Preempt amortizes transition delays and improves aggregate throughput under bursty or imbalanced workloads. This makes it well-suited for dynamically rebalancing between "many DP engines" and "few fast TP engines" as load fluctuates.

\subsubsection{Hard Preempt (Latency-Oriented)}

\textit{Hard Preempt} targets scenarios where immediate service is required, aligning directly with priority-aware service differentiation in Use Case~2. When a high-priority TP request arrives, the scheduler interrupts all active DP execution across participating engines and executes the TP request without waiting for stragglers (Figure~\ref{fig:switching_strategies}(c)).

Crucially, this interruption does not impose recomputation overhead on paused DP requests. Leveraging the unified KV Cache Adaptor (Section~\ref{sec:kv_adaptor}), KV blocks with different logical layouts can coexist in the same physical memory pool. As a result, interrupted DP workloads retain valid KV state and resume execution seamlessly after the TP request completes. Hard Preempt thus enforces low tail latency without sacrificing correctness or throughput for background traffic.


\subsection{Discussion}

\subsubsection{Generality.}
While motivated by specific user scenarios, Soft and Hard Preempt represent general scheduling primitives rather than workload-specific heuristics. Soft Preempt captures a broad class of speculative execution techniques that trade limited recomputation for improved utilization, while Hard Preempt generalizes preemptive scheduling to distributed TP execution without KV migration. Together, they allow the scheduler to continuously navigate the latency vs. throughput vs. memory trade-off space under dynamic conditions.

Importantly, both strategies extend the base execution flow without violating its core invariants: TP execution observes a globally consistent request order, and all mode transitions occur only at scheduler-coordinated safe points. This separation of \emph{mechanism} (dynamic DP--TP switching) from \emph{policy} (Soft vs.\ Hard Preempt) enables \fram to support a wide range of serving objectives without redesigning the underlying system.

\subsubsection{Limitations.} 

While \fram's design focuses on improving throughput and latency flexibility, the current implementation is designed primarily for intra-node parallelism. Specifically, our system assumes that the model weights and the necessary KV cache can be managed within the memory hierarchy of a single multi-GPU node (e.g., 8$\times$ H200). Consequently, extremely large models that necessitate multi-node model parallelism are outside the current scope of this work. However, we note that the vast majority of popular open-source models (e.g., Llama-3-70B, GPT-oss-120B) fit comfortably within a single modern GPU node, allowing multiple DP instances to run concurrently. Extending our dynamic reconfiguration protocol to support inter-node communication remains a promising direction for future research.
\section{Evaluation}

\subsection{Experimental Setup}
\label{sec:setup}

\subsubsection{Hardware}

We conduct our experiments on the \textsc{Defiant} cluster at Oak Ridge National Laboratory (OLCF ACE testbed)~\cite{olcf_defiant}. The system consists of 20 HPE Cray XD220 CPU nodes along with 2 HPE Cray XD670 Nvidia H200 GPU nodes, and each node is equipped with 8$\times$ NVIDIA H200 GPUs. Each H200 GPU provides 141~GB of HBM3e (4.8~TB/s peak bandwidth) and 1{,}979~TFLOPS peak dense FP8 throughput. The 8 GPUs are interconnected via NVLink, providing 900~GB/s bi-directional bandwidth.


\subsubsection{Software, LLM Serving Models and Baselines} We implement \fram on top of vLLM v1. 
We evaluate \fram on three representative models from today's widely used LLM families: LLama, GPT, and Nemotron, spanning dense, sparse, and long-context serving:
\begin{itemize}[leftmargin=*]
    \item \textbf{Llama-3 (70B, dense)}~\cite{llama3_herd}: a large, dense Transformer baseline that stresses compute and all-reduce bandwidth without sparsity.
    \item \textbf{GPT-OSS (120B, MoE)}~\cite{gptoss_modelcard}: a sparse Mixture-of-Experts model that activates only a small subset of experts per token, stressing routing, load balance, and sparse execution.
    \item \textbf{Nemotron (8B, dense)}~\cite{xu2025ultralong}: an ultra-long context model that can digest up to 4M tokens to stress KV cache capacity and memory pressure.
\end{itemize}

We compare against three baselines: (i) \textbf{Static DP}, (ii) \textbf{Static TP}, and (iii) \textbf{Shift-Parallelism}~\cite{shift_parallelism}. Shift-Parallelism is a SoTA baseline as a \emph{production-deployed} vLLM-integrated system that enables seamless runtime switching between latency-optimal TP and throughput-oriented sequence parallelism (SP) by exploiting KV cache invariance. The prior results show that this design has a strictly better latency-throughput trade-off than static DP/TP across arrival rates (e.g., up to $\sim$50\% higher throughput and $\sim$1.51$\times$ faster response, with lowest completion time over rates), making it the most direct comparison point for our dynamic DP-TP switching.

\subsubsection{Datasets and Synthetic Workloads}

We evaluate on three open-sourced datasets: ShareGPT~\cite{hu2023chatgpt}, CodeActInstruct~\cite{wang2024executable}, and HumanEval~\cite{chen2021evaluating}, covering conversational chat-based assistance, code-centric instruction following, and program synthesis.

We developed a tool to generate synthetic workloads to mimic real-world user requests because publicly available LLM datasets provide request contents but not realistic, reproducible arrival-time traces. A synthesized workload lets us precisely vary burst intensity and duration to stress mode-switching behavior and make comparisons repeatable across systems. This design aligns with the evaluation method from \cite{shift_parallelism}.
Specifically, we synthesize
(1) \textit{request lengths} with prompts sampled uniformly span $[128, 4000]$ input tokens and $[64, 512]$ output tokens;
(2) \textit{traffic pattern} where the arrival rate alternates between low load (2--5 req/s) and high load bursts (10--30 req/s);
(3) \textit{volume} in which each iteration issues 4000 requests to capture steady-state behavior across multiple bursts.

\subsubsection{Performance Metrics}
We use standard streaming-inference metrics that quantify initial responsiveness and steady-state token generation~\cite{agrawal2024taming}:
%

(i) \textbf{Time To First Token (TTFT):} latency from when a request arrives at the serving system to when the first output token is generated (including both queuing and prefill). 

(ii) \textbf{Time Per Output Token (TPOT):} the per-request average \emph{time-between-tokens} during decoding, measured over consecutive output tokens after the first (i.e., inter-token interval).

(iii)  \textbf{Peak generation throughput:}  maximum aggregate output token rate (tokens/s) sustained by the system under load.

(iv) \textbf{Queue time:} time from request admission to first scheduling, isolating scheduler delay from execution time~\cite{vllm_metrics}.


We export per-request and system metrics from the serving engine to Prometheus\cite{prometheus_io} and visualize time-series behavior in Grafana\cite{elradi2025prometheus}. To minimize instrumentation overhead on the critical path, we log concurrency, TTFT, and queue time in the backend, and compute TPOT and aggregate throughput at the client from token timestamps.

\begin{figure*}[t]
    \centering
    \begin{subfigure}{0.32\textwidth}
        \includegraphics[width=\linewidth]{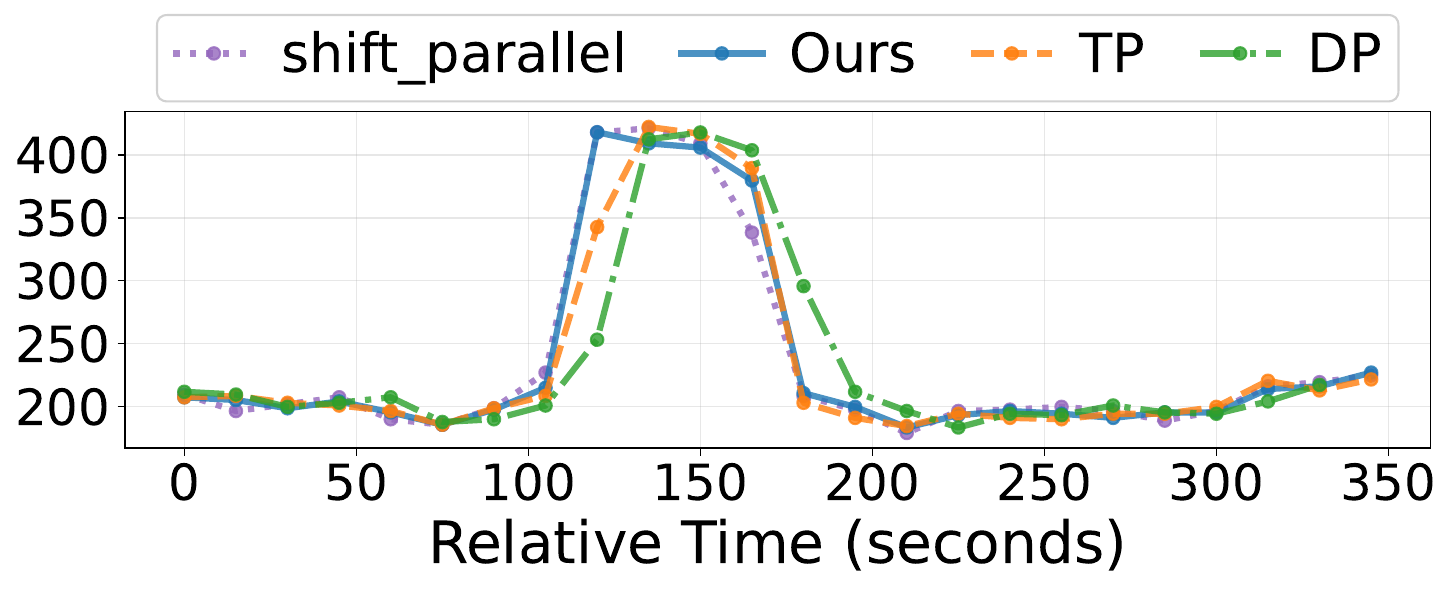}
        \caption{Llama-70B: Concurrency}
        \label{fig:req_70b}
    \end{subfigure}
    \hfill
    \begin{subfigure}{0.32\textwidth}
        \includegraphics[width=\linewidth]{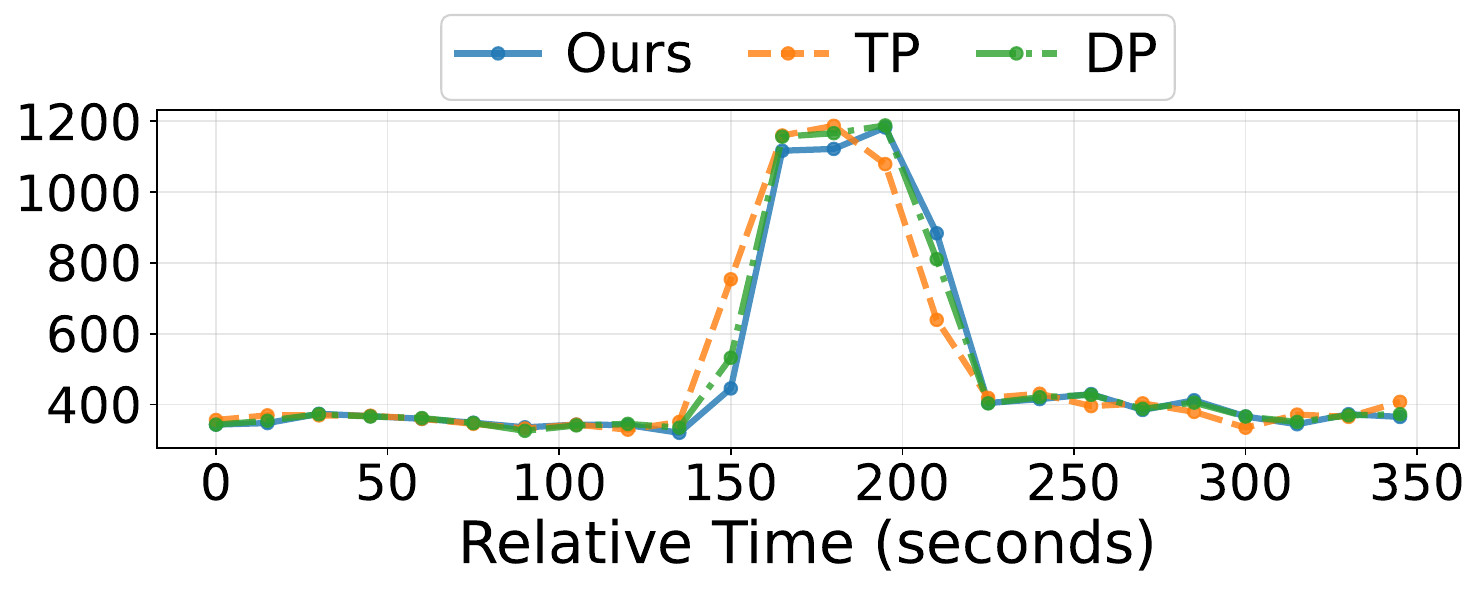} 
        \caption{GPT-OSS-120B: Concurrency}
        \label{fig:req_gpt}
    \end{subfigure}
    \hfill
    \begin{subfigure}{0.32\textwidth}
        \includegraphics[width=\linewidth]{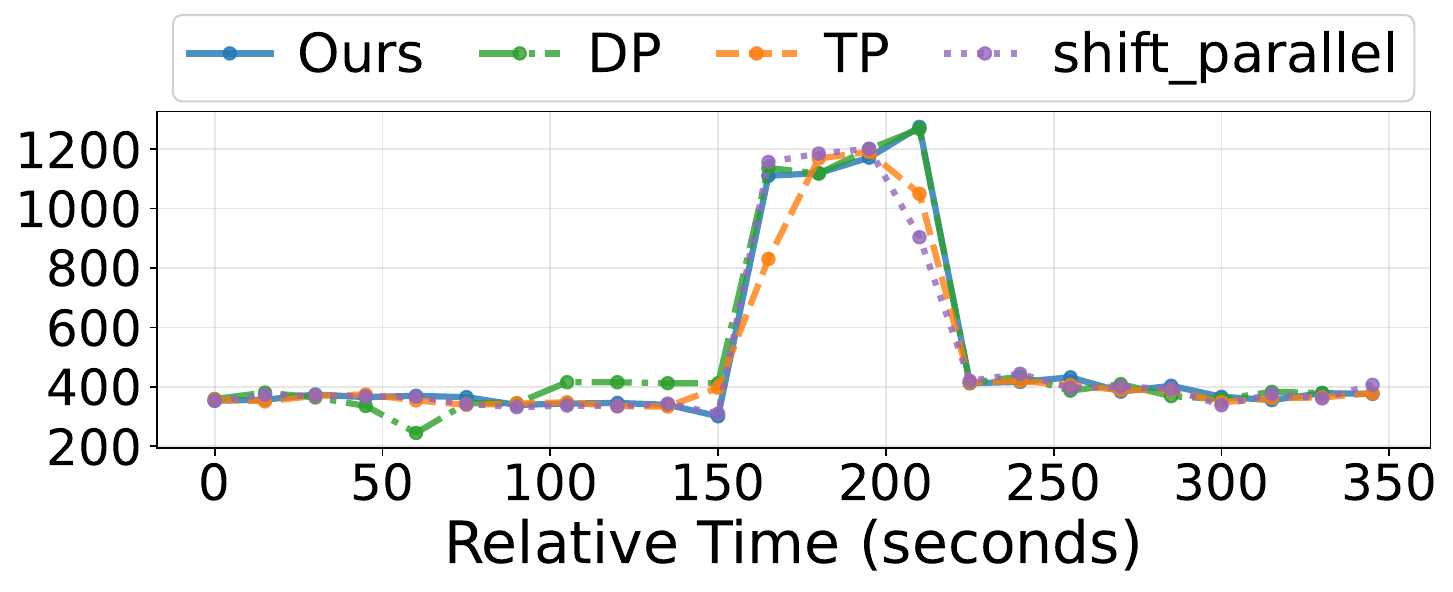} 
        \caption{Nemotron-8B: Concurrency}
        \label{fig:req_8b}
    \end{subfigure}
    
    \vspace{0.1cm} 
    
    \begin{subfigure}{0.32\textwidth}
        \includegraphics[width=\linewidth]{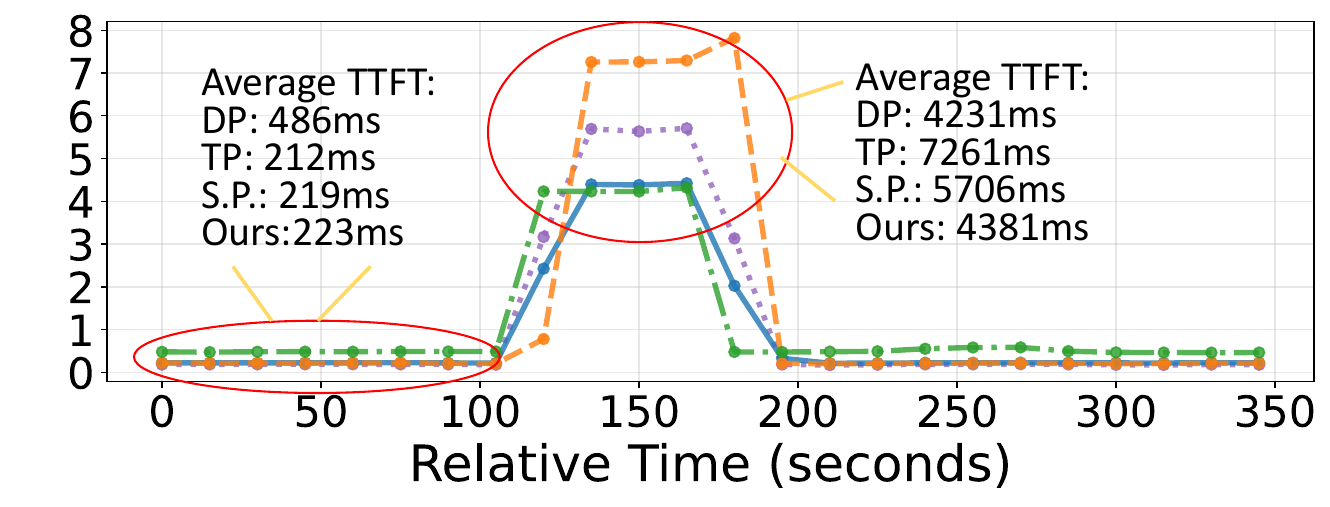}
        \caption{Llama-70B: TTFT (P90)}
        \label{fig:ttft_70b}
    \end{subfigure}
    \hfill
    \begin{subfigure}{0.32\textwidth}
        \includegraphics[width=\linewidth]{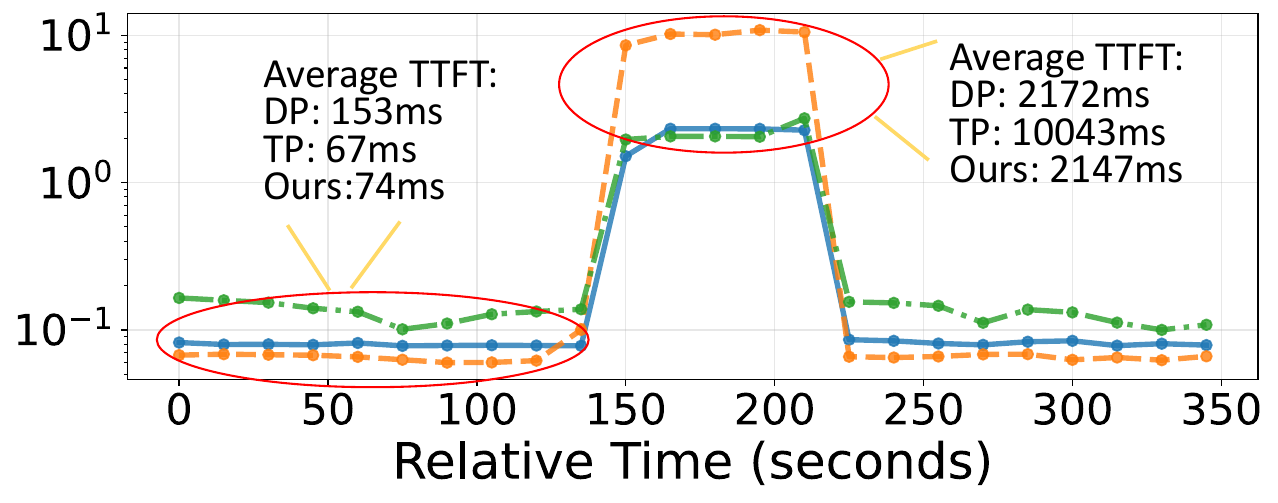} 
        \caption{GPT-OSS-120B: TTFT (P90)}
        \label{fig:ttft_gpt}
    \end{subfigure}
    \hfill
    \begin{subfigure}{0.32\textwidth}
        \includegraphics[width=\linewidth]{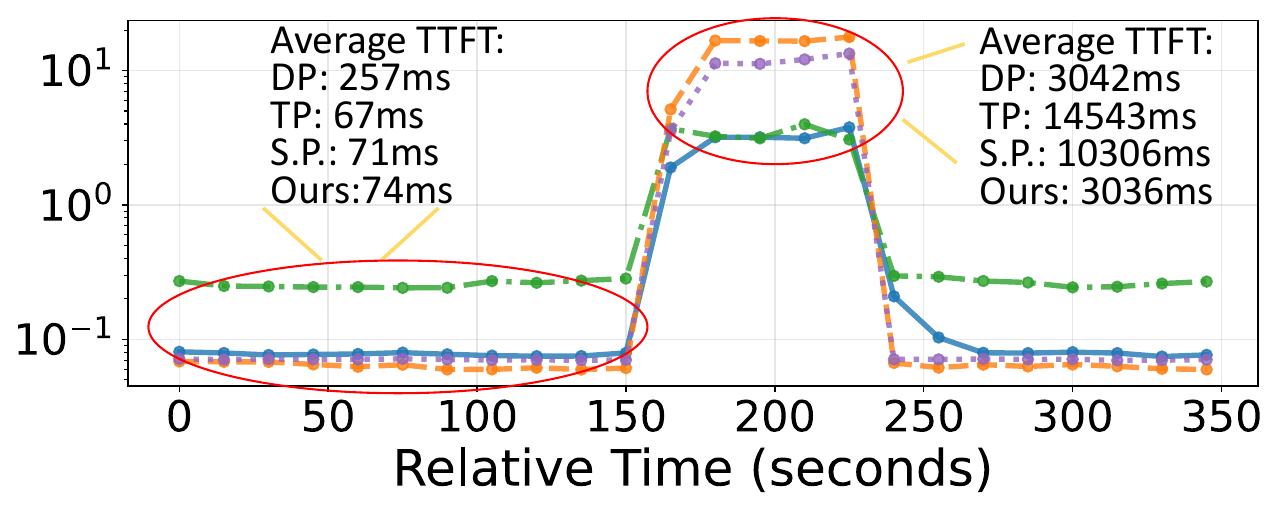} 
        \caption{Nemotron-8B: TTFT (P90)}
        \label{fig:ttft_8b}
    \end{subfigure}

    \vspace{0.1cm} 

    \begin{subfigure}{0.32\textwidth}
        \includegraphics[width=\linewidth]{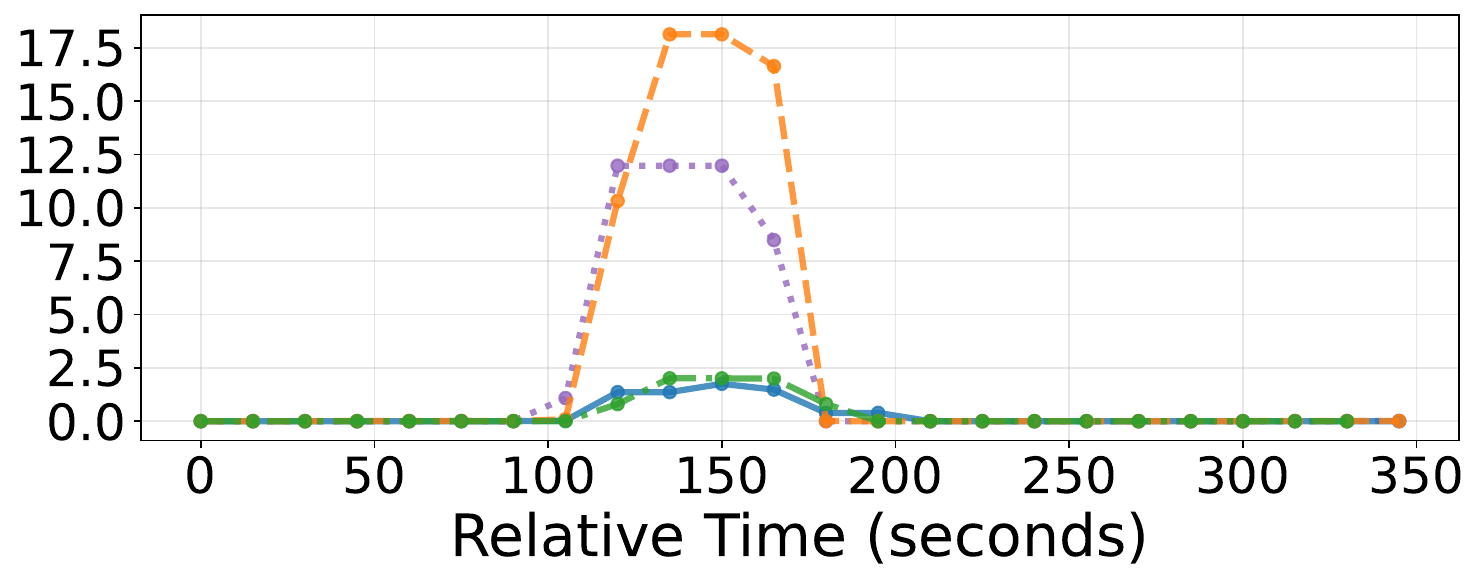}
        \caption{Llama-70B: Queue Time}
        \label{fig:qtime_70b}
    \end{subfigure}
    \hfill
    \begin{subfigure}{0.32\textwidth}
        \includegraphics[width=\linewidth]{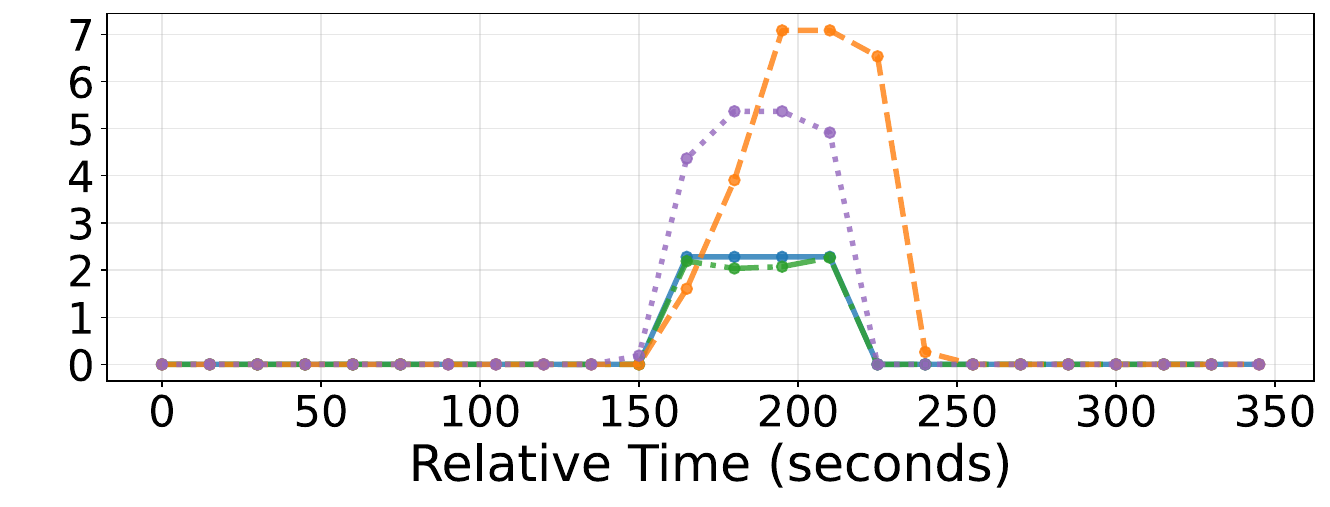} 
        \caption{GPT-OSS-120B: Queue Time}
        \label{fig:qtime_gpt}
    \end{subfigure}
    \hfill
    \begin{subfigure}{0.32\textwidth}
        \includegraphics[width=\linewidth]{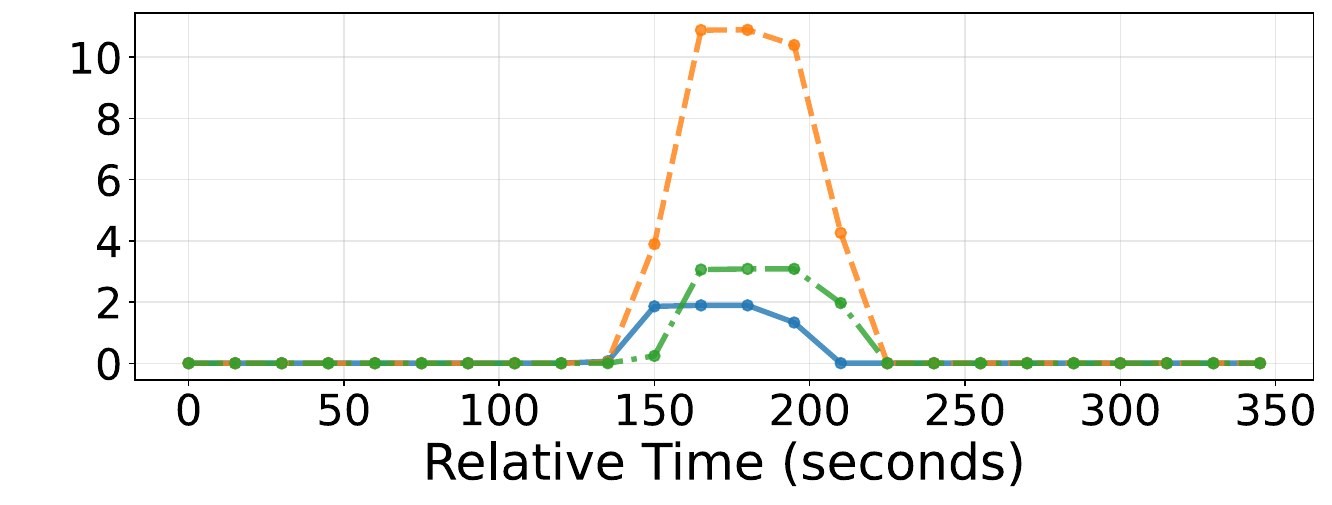} 
        \caption{Nemotron-8B: Queue Time}
        \label{fig:qtime_8b}
        
    \end{subfigure}
\caption{\textbf{End-to-end performance under bursty traffic.} Columns show Llama-3-70B (left), GPT-OSS-120B (middle), and Nemotron-8B (right). Rows report in-flight concurrency, P90 TTFT, and queue time. \fram (blue) tracks load shifts: it avoids the TTFT/queue spikes of static TP (orange) during bursts, stays close to TP at low load, and substantially improves over static DP (green). Where supported, \fram also outperforms the SoTA dynamic baseline Shift-Parallelism (purple).}

    \Description{A 3x3 grid of line charts.}
    \label{fig:overall_performance_all_models}

\end{figure*}

\subsection{Overall Performance}
\label{sec:overall_eval}

\begin{figure}[ht]
  \centering
  \begin{subfigure}[t]{0.49\linewidth}
    \centering
    \includegraphics[width=\linewidth]{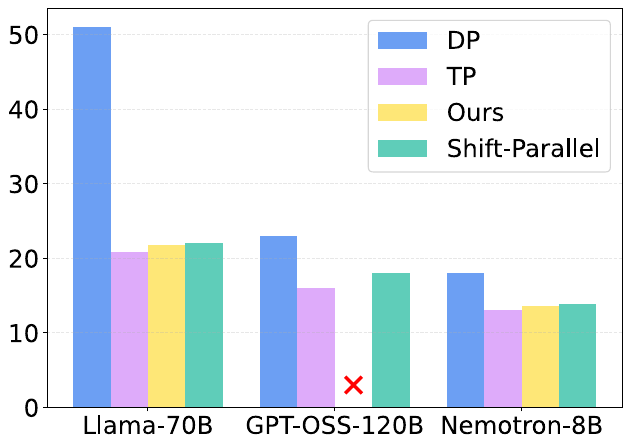}
    \caption{Median TPOT (ms)}
    \label{fig:req_70b}
  \end{subfigure}
  \hfill 
  \begin{subfigure}[t]{0.49\linewidth}
    \centering
    \includegraphics[width=\linewidth]{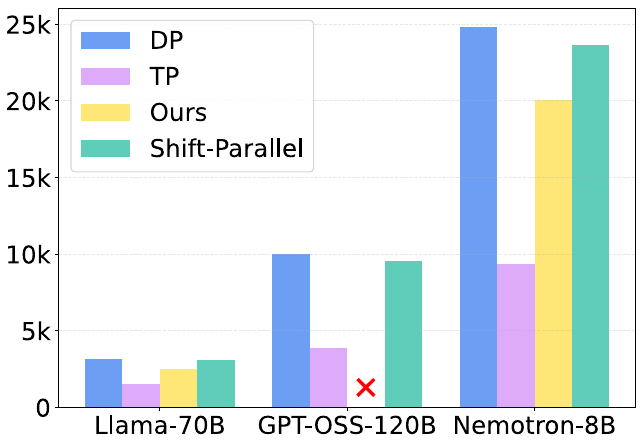}
    \caption{Peak Throughput (tokens/s)}
    \label{fig:req_gpt}
  \end{subfigure}
  
  \caption{Comparison of median TPOT and peak generation throughput across different models.}
  \vspace{-2em}
  \Description{Comparison of median TPOT and peak generation throughput.}
  \label{fig:topt_throughput}
\end{figure}

We compare \fram against static TP, static DP, and Shift-Parallelism using the bursty workload from Section~\ref{sec:setup}\footnote{Shift-Parallelism does not yet support GPT-OSS-120B due to compatibility issues with the new MoE kernel, so we are currently unable to report its results on GPT-OSS.}. Figure~\ref{fig:overall_performance_all_models} plots (top) in-flight concurrency, (middle) P90 TTFT, and (bottom) queue time over the trace; Figure~\ref{fig:topt_throughput} summarizes steady-state TPOT and peak generation throughput.




\textit{Same offered load.}
All systems observe the same concurrency pattern (Figure~\ref{fig:overall_performance_all_models}(a--c)), ensuring the latency (Figure~\ref{fig:overall_performance_all_models}(d--f)) and queueing differences (Figure~\ref{fig:overall_performance_all_models}(g--i)) are due to execution strategy rather than input load.


\textit{Bursts: \fram matches DP by avoiding queue build-up.}
During high-load phases, static TP (and Shift-Parallelism where applicable) accumulates substantial queueing (Figure~\ref{fig:overall_performance_all_models}(g--i)), which dominates TTFT (Figure~\ref{fig:overall_performance_all_models}(d--f)). In contrast, \fram switches to DP and keeps queue time near DP, yielding TTFT close to the DP lower bound (e.g., average TTFT at burst: Llama-70B: 4.38s vs.\ 4.23s DP; GPT-OSS-120B: 2.15s vs.\ 2.17s DP; Nemotron-8B: 3.04s vs.\ 3.04s DP). Relative to static TP, \fram reduces P90 TTFT by $1.66\times$ (Llama-70B), $4.68\times$ (GPT-OSS-120B), and $4.79\times$ (Nemotron-8B). Relative to Shift-Parallelism, \fram reduces P90 TTFT by $1.3\times$ (Llama-70B) and $3.39\times$ (Nemotron-8B).



\textit{Light loads (flat periods): \fram stays near TP with small overhead.}
Under low load, \fram remains in TP to minimize TTFT, closely tracking static TP (e.g., average TTFT: Llama-70B: 223ms vs.\ 212ms TP; GPT-OSS-120B: 74ms vs.\ 67ms TP; Nemotron-8B: 74ms vs.\ 67ms TP) while substantially improving over static DP. The remaining gap to TP reflects only mode-management overhead: $5.19\%$ (12ms) for Llama-70B and $10.45\%$ (7ms) for GPT-OSS-120B and $10.45\%$ (7ms) for Nemotron-8B. In stead, Shift-Parallelism is up to $\sim$4\% faster at low load, but incurs significantly higher burst latency (3.39x than ours) and queueing (2.48x than ours).

\textit{TPOT and throughput: near-TP latency with near-DP capacity.}
Figure~\ref{fig:topt_throughput} shows that \fram improves median TPOT over static DP by $2.31\times$, $1.28\times$, and $1.30\times$ for Llama-70B, GPT-OSS-120B, and Nemotron-8B, respectively, approaching TP-like per-token latency. At the same time, \fram retains $\approx$95--96\% of DP peak throughput across models (e.g., 3,059 vs.\ 3,169 tokens/s on the largest size model Llama-70B), and outperforms static TP by $2.03\times$, $2.47\times$, and $2.52\times$ in peak throughput. Where supported, \fram also exceeds Shift-Parallelism in peak throughput (1.22$\times$ on Llama-70B; 1.18$\times$ on Nemotron-8B).

\subsection{Performance on Workloads with Priority}
\label{sec:piority}


We evaluate how \fram handles mixed-priority serving on Llama-70B. The workload interleaves high-priority and normal requests, while the arrival rate is modulated between 3-5 requests/s to create sustained queueing pressure. We compare against two static baselines, i.e., static TP and static DP, under the same workload and report mean TPOT/TTFT (for priority requests and for all requests) plus peak throughput, following standard comparative methodology in systems evaluations. Table~\ref{tab:priority_performance} summarizes the results.


\begin{table}[ht]
\centering
\caption{Llama-70B under mixed-priority workload.}
\vspace{-10pt}
\label{tab:priority_performance}
\begin{tabular}{l|ccc}
\toprule
\textbf{Metric} & \textbf{static TP} & \textbf{static DP} & \textbf{Ours} \\
\midrule
Mean TPOT (priority) (ms) & 22 & 32 & 24 \\
Mean TPOT (all) (ms) & 22 & 32 & 28 \\
Mean TTFT (priority) (ms) & 63 & 166 & 74 \\
Mean TTFT (all) (ms) & 2130 & 166 & 142 \\
Peak Throughput (tokens/s) & 2530 & 3164 & 3040 \\
\bottomrule
\end{tabular}
\end{table}




\textit{Priority requests.}
\fram preserves near-TP latency for high-priority traffic: mean TPOT is 24\,ms and mean TTFT is 74\,ms, within 1.09$\times$ and 1.17$\times$ of the static TP baseline (22\,ms / 63\,ms). Relative to static DP, \fram improves priority TPOT by 1.33$\times$ (32$\rightarrow$24\,ms) and TTFT by 2.24$\times$ (166$\rightarrow$74\,ms).

\textit{Overall system behavior under load.}
Static TP becomes throughput-limited and suffers severe queue time: mean TTFT (all requests) increases to 2130\,ms. \fram avoids this collapse by adapting parallelism to drain backlog, reducing mean TTFT (all) to 142\,ms (15.0$\times$ lower than TP), and remaining 1.17$\times$ better than static DP (166\,ms).

\textit{Throughput.}
\fram sustains 3040 tokens/s peak throughput, retaining 96\% of the DP baseline (3164 tokens/s) while delivering TP-like latency for priority requests.

\subsection{Max Context Length and Switching Latency}
\label{sec:cont-length}

We quantify \fram's ability to (i) expand KV cache capacity for long-context requests and (ii) reconfigure parallelism online. Unless otherwise noted, experiments run on 8$\times$ NVIDIA H200 GPUs hosting Llama-3-70B.

\subsubsection{Context-Length Capacity}


The maximum context length is bounded by the KV cache memory left after loading model weights. Table~\ref{tab:context_switch} shows that static deployments impose rigid limits tied to the fixed TP degree: $4\text{DP}\!\times\!2\text{TP}$ can only support 264K tokens, and $2\text{DP}\!\times\!4\text{TP}$ only supports 959K tokens.
By dynamically merging workers, \fram scales KV capacity on demand and supports up to 1.9M tokens, a $7.2\times$ increase over $4\text{DP}\!\times\!2\text{TP}$, a $2.0\times$ increase over $2\text{DP}\!\times\!4\text{TP}$, and within 17\% of the static $1\text{DP}\!\times\!8\text{TP}$ (the upper bound - "2.3M"). This result evaluates that \fram's reconfiguration support for dynamic DP-TP shifting introduces a small, fixed memory consumption; however, it frees up nearly all remaining GPU memory for the KV cache.

\begin{table}[h]
    \centering
    \caption{Max context support and switching latency.}
    \vspace{-10pt}
    \label{tab:context_switch}
    \resizebox{\linewidth}{!}{
        \begin{tabular}{l c c c}
        \toprule
        \textbf{Configuration} & \textbf{GPUs/Inst.} & \textbf{Max Context} & \textbf{Switching Latency} \\
        \midrule
        Static $4\text{DP} \times 2\text{TP}$ & 2 & 264 K & 292.38 s (Cold Start) \\
        Static $2\text{DP} \times 4\text{TP}$ & 4 & 959 K & 211.97 s (Cold Start) \\
        Static $1\text{DP} \times 8\text{TP}$ & 8 & 2.3 M & 146.54 s (Cold Start) \\
        \fram & dynamic & 1.9 M & 15 ms (Live) \\
        \bottomrule
        \end{tabular}%
    }
\end{table}

\subsubsection{Switching Latency}


Static systems handle an over-limit request via a cold restart (shut down and re-launch with a higher TP degree), which requires reloading weights and re-initializing collectives; Table~\ref{tab:context_switch} shows these static methods taking non-negligible overhead, i.e., 146-292 seconds.
In contrast, \fram performs live switching using the pre-initialized \textit{Communicator Pool} and zero-copy \textit{Model Weights Manager} in Section~\ref{sec:system_design}, completing the switch in 15\,ms, about five orders of magnitude faster ($\sim$10{,}000$\times$), making dynamic DP-TP switching practical for large-scale, latency-critical serving.

\begin{figure}[ht]
  \centering
  \includegraphics[width=\linewidth]{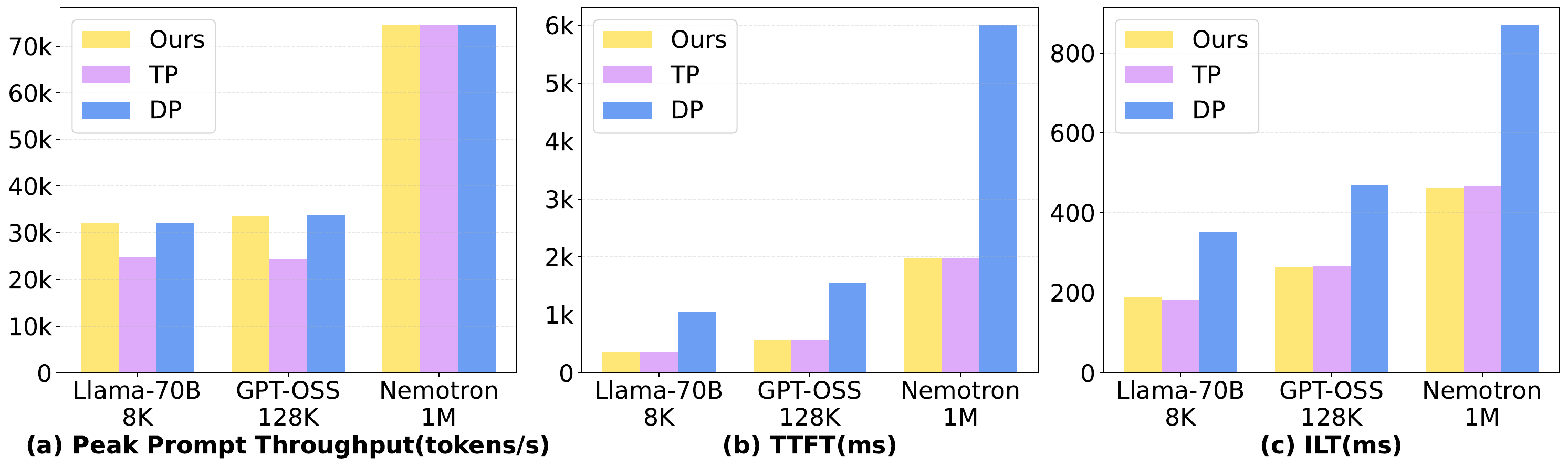}
    \caption{Comparison of Peak Prompt Throughput, TTFT, and ILT under long-contexts (8K, 128K, and 1M).}
  \Description{topt and throughput.}
  \label{fig:stress_test}
\end{figure}



\subsection{Ultra Long-Context Stress Test}


We further stress the system at each model's maximum supported context length (8K for Llama-70B, 128K for GPT-OSS-120B, and 1M for Nemotron). Figure~\ref{fig:stress_test} reports peak prompt throughput, TTFT, and ILT\footnote{Inter-Token Latency (ILT) measures the average time interval between the generation of consecutive tokens. We report ILT (instead of TPOT) because TPOT aggregates both compute and queueing/batching effects and can vary with scheduler decisions}

\textit{Peak Prompt Throughput.}
%
Figure~\ref{fig:stress_test} (a) shows that \fram sustains DP-level peak prompt throughput at the maximum context across all models. This translates to $1.29\times$ (Llama-70B)and $1.38\times$  (GPT-OSS-120B)higher throughput than static TP. For Nemotron-8B(1M), all three configurations converge at approximately 74K tokens/s, as the extremely long context relies more on memory bandwidth.

\textit{Latency Performance (TTFT and ILT).}
Figure~\ref{fig:stress_test} (b) shows our system achieves the TTFT performance nearly to the TP-like latency, with small overheads ($<1.1\%$). Also, compared to the static DP baseline, \fram delivers substantial speedups of $2.94\times$, $2.78\times$, and $3.04\times$ for Llama-70B, GPT-OSS-120B, and Nemotron-8B, respectively.
%
Figure~\ref{fig:stress_test}(c) shows that our system maintains TP-like inter-token latency, staying within 5\% of static TP, while reducing ILT by $1.85-1.88\times$ relative to static DP across all models.


Overall, \fram removes the need to pre-commit to a single static configuration: it preserves DP-like capacity for long-context bursts while maintaining TP-like latency when load is light.

\section{Related Works}



Recent research in distributed LLM inference has shifted from static configurations to dynamic adaptability. 

\textbf{Dynamic Parallelism Transformation.}
Several systems enable runtime adaptability but face specific limitations. 
Hetis~\cite{mo2025hetis} uses a fine-grained and dynamic parallelism architecture that enables precise control over heterogeneous resources.
LoongServe~\cite{wu2024loongserve}.
Gyges~\cite{chen2025gyges} accelerates state redistribution via weight padding yet still incurs latency penalties in the range of hundreds of milliseconds. Shift Parallelism~\cite{hidayetoglu2025shift} exploits KV cache invariance for seamless transitions between Tensor and Sequence Parallelism but is restricted to specific hardware topologies. Seesaw~\cite{seesaw2025} dynamically re-shards models between prefill and decode phases; however, the overhead of frequent reconfiguration can negate throughput gains. Similarly, HAP~\cite{liu2025hap} employs Integer Linear Programming to optimize hybrid strategies for MoE models, but the solver introduces significant runtime complexity.

\textbf{Static and Scheduling-Based Optimization.}
Other approaches prioritize static planning or state abstraction. AlpaServe~\cite{li2023alpaserve} uses compilation to generate optimal static plans, but cannot adapt running instances to traffic bursts without restarting. Tenplex~\cite{wagenlander2024tenplex} abstracts model state for dynamic training, yet lacks the sub-second responsiveness required for interactive inference. Finally, FastServe~\cite{wu2023fastserve} optimizes throughput via preemptive scheduling and host memory offloading, but relies on limited PCIe bandwidth rather than flexible parallelism reconfiguration.



\section{Conclusion}
\label{sec:conclusion}
\fram shows that static parallelism not be a deployment-time choice for LLM serving. By virtualizing weights, KV state, and communication, \fram enables online DP$\leftrightarrow$TP reconfiguration as a lightweight operation coordinated by a deadlock-free scheduler. Across three models and realistic workloads, this capability translates into up to $4.79\times$ speedup at high load and $3.47\times$ at low load, while accommodating priority and long-context requests.

\section{Acknowledgment}

This work is supported by the U.S. National Science Foundation (2514351 and 2505118) and by the Graduate Research at ORNL (GRO) Internship Program at Oak Ridge National Laboratory. This research used resources of the Oak Ridge Leadership Computing Facility (OLCF), which is a DOE Office of Science User Facility at the Oak Ridge National Laboratory supported by the U.S. Department of Energy under Contract No. DE-AC05-00OR22725. We thank the anonymous reviewers for their valuable feedback.


\bibliographystyle{unsrt}
\bibliography{sample-base}

@misc{nvidia2024perplexity,
  title        = {Spotlight: Perplexity AI Serves 400 Million Search Queries a Month Using {NVIDIA} Inference Stack},
  author       = {Elmeleegy, Amr and Chen, Lequn and Hu, Kevin},
  year         = {2024},
  month        = dec,
  howpublished = {\url{https://developer.nvidia.com/blog/spotlight-perplexity-ai-serves-400-million-search-queries-a-month-using-nvidia-inference-stack/}}
}

@article{huang2019gpipe,
  title={Gpipe: Efficient training of giant neural networks using pipeline parallelism},
  author={Huang, Yanping and Cheng, Youlong and Bapna, Ankur and Firat, Orhan and Chen, Dehao and Chen, Mia and Lee, HyoukJoong and Ngiam, Jiquan and Le, Quoc V and Wu, Yonghui and others},
  journal={Advances in neural information processing systems},
  volume={32},
  year={2019}
}

@article{liu2024deepseek,
  title={Deepseek-v3 technical report},
  author={Liu, Aixin and Feng, Bei and Xue, Bing and Wang, Bingxuan and Wu, Bochao and Lu, Chengda and Zhao, Chenggang and Deng, Chengqi and Zhang, Chenyu and Ruan, Chong and others},
  journal={arXiv preprint arXiv:2412.19437},
  year={2024}
}

@misc{jaiswal2025sageserve,
  title         = {Serving Models, Fast and Slow: Optimizing Heterogeneous {LLM} Inferencing Workloads at Scale},
  author        = {Jaiswal, Shashwat and Jain, Kunal and Simmhan, Yogesh and Parayil, Anjaly and Mallick, Ankur and Wang, Rujia and St. Amant, Renee and Bansal, Chetan and R{\"u}hle, Victor and Kulkarni, Anoop and Kofsky, Steve and Rajmohan, Saravan},
  year          = {2025},
  eprint        = {2502.14617},
  archivePrefix = {arXiv},
  primaryClass  = {cs.DC},
  doi           = {10.48550/arXiv.2502.14617}
}

@misc{wang2024burstgpt,
  title         = {BurstGPT: A Real-world Workload Dataset to Optimize {LLM} Serving Systems},
  author        = {Wang, Yuxin and Chen, Yuhan and Li, Zeyu and Kang, Xueze and Tang, Zhenheng and He, Xin and Guo, Rui and Wang, Xin and Wang, Qiang and Zhou, Amelie Chi and Chu, Xiaowen},
  year          = {2024},
  eprint        = {2401.17644},
  archivePrefix = {arXiv},
  primaryClass  = {cs.DC},
  doi           = {10.48550/arXiv.2401.17644}
}

@misc{azure2024opt_infer,
  title        = {Optimizing Language Model Inference on {Azure}},
  author       = {Patankar, Shantanu Deepak and Affaticati, Hugo},
  year         = {2024},
  month        = oct,
  howpublished = {\url{https://techcommunity.microsoft.com/blog/azurehighperformancecomputingblog/optimizing-language-model-inference-on-azure/4248271}}
}

@misc{dynamollm2024,
  title         = {DynamoLLM: Designing {LLM} Inference Clusters for Performance and Energy Efficiency},
  author        = {Stojkovic, Jovan and Zhang, Chaojie and Goiri, {\'I}{\~n}igo and Torrellas, Josep and Choukse, Esha},
  year          = {2024},
  eprint        = {2408.00741},
  archivePrefix = {arXiv},
  primaryClass  = {cs.DC},
  doi           = {10.48550/arXiv.2408.00741}
}

@inproceedings{reddi2019mlperf,
  title     = {{MLPerf} Inference Benchmark},
  author    = {Reddi, Vijay Janapa and Cheng, Christine and Kanter, David and Mattson, Peter and Schmuelling, Guenther and Wu, Carole-Jean and Anderson, Brian and Breughe, Maximilien and Charlebois, Mark and Chou, William and others},
  booktitle = {Proceedings of the 47th Annual International Symposium on Computer Architecture (ISCA)},
  year      = {2020},
  doi       = {10.1109/ISCA45697.2020.00045},
  note      = {arXiv:1911.02549}
}

@misc{mlperf_inference_docs,
  title        = {{MLPerf} Inference Benchmarks Documentation},
  author       = {{MLCommons}},
  year         = {2025},
  howpublished = {\url{https://docs.mlcommons.org/inference/}}
}

@misc{mlcommons2025llm_v5,
  title        = {{MLPerf} Inference v5.0 Advances Language Model Capabilities for {GenAI}},
  author       = {{MLCommons}},
  year         = {2025},
  month        = apr,
  howpublished = {\url{https://mlcommons.org/2025/04/llm-inference-v5/}}
}

@misc{nvidia2025llm_benchmarking,
  title        = {{LLM} Inference Benchmarking: Fundamental Concepts},
  author       = {Nguyen, Vinh and Gao, Wenwen and Apsey, Emily and Kudleppanavar, Ganesh and Shah, Neelay and Bermudez, Elias},
  year         = {2025},
  month        = apr,
  howpublished = {\url{https://developer.nvidia.com/blog/llm-benchmarking-fundamental-concepts/}}
}

@misc{lai2025tokenscale,
  title         = {TokenScale: Timely and Accurate Autoscaling for Disaggregated LLM Serving with Token Velocity},
  author        = {Ruiqi Lai and Hongrui Liu and Chengzhi Lu and Zonghao Liu and Siyu Cao and Siyang Shao and Yixin Zhang and Luo Mai and Dmitrii Ustiugov},
  year          = {2025},
  eprint        = {2512.03416},
  archivePrefix = {arXiv},
  primaryClass  = {cs.DC},
  doi           = {10.48550/arXiv.2512.03416}
}

@misc{stojkovic2024dynamollm,
  title         = {DynamoLLM: Designing LLM Inference Clusters for Performance and Energy Efficiency},
  author        = {Jovan Stojkovic and Chaojie Zhang and {\'I}{\~n}igo Goiri and Josep Torrellas and Esha Choukse},
  year          = {2024},
  eprint        = {2408.00741},
  archivePrefix = {arXiv},
  primaryClass  = {cs.DC},
  doi           = {10.48550/arXiv.2408.00741}
}

@misc{liu2025greenllm,
  title         = {GreenLLM: SLO-Aware Dynamic Frequency Scaling for Energy-Efficient LLM Serving},
  author        = {Qunyou Liu and Darong Huang and Marina Zapater and David Atienza},
  year          = {2025},
  eprint        = {2508.16449},
  archivePrefix = {arXiv},
  primaryClass  = {cs.PF},
  doi           = {10.48550/arXiv.2508.16449}
}

@misc{li2025adaserve,
  title         = {AdaServe: SLO-Customized LLM Serving with Fine-Grained Speculative Decoding},
  author        = {Zikun Li and Zhuofu Chen and Remi Delacourt and Gabriele Oliaro and Zeyu Wang and Qinghan Chen and Shuhuai Lin and April Yang and Zhihao Zhang and Zhuoming Chen and Sean Lai and Xupeng Miao and Zhihao Jia},
  year          = {2025},
  eprint        = {2501.12162},
  archivePrefix = {arXiv},
  primaryClass  = {cs.DC},
  doi           = {10.48550/arXiv.2501.12162}
}

@misc{nguyen2025llm_benchmarking,
  title        = {{LLM} Inference Benchmarking: Fundamental Concepts},
  author       = {Nguyen, Vinh and Gao, Wenwen and Apsey, Emily and Kudleppanavar, Ganesh and Shah, Neelay and Bermudez, Elias},
  year         = {2025},
  month        = apr,
  howpublished = {\url{https://developer.nvidia.com/blog/llm-benchmarking-fundamental-concepts/}}
}

@misc{vllm-linear-doc,
  author       = {{vLLM Project}},
  title        = {{vLLM API Reference: vllm.model\_executor.layers.linear (Source: vllm/model\_executor/layers/linear.py)}},
  howpublished = {\url{https://docs.vllm.ai/en/v0.10.1/api/vllm/model_executor/layers/linear.html}},
  note         = {}
}

@misc{vllm-v1,
  author       = {{vLLM Project}},
  title        = {{vLLM v1: Next-Generation LLM Serving Engine}},
  howpublished = {\url{https://docs.vllm.ai/en/latest/}},
  note         = {Released January 2025}
}

@misc{olcf_defiant,
  title        = {Defiant Quick-Start Guide},
  author       = {{Oak Ridge Leadership Computing Facility (OLCF)}},
  howpublished = {\url{https://docs.olcf.ornl.gov/ace_testbed/defiant_quick_start_guide.html}},
  note         = {}
}

@article{rajbhandari2025arctic,
  title={Arctic Inference with Shift Parallelism: Fast and Efficient Open Source Inference System for Enterprise AI},
  author={Rajbhandari, Samyam and Hidayetoglu, Mert and Qiao, Aurick and Wang, Ye and Yang, Juncheng and Rasley, Jeff and Wyatt, Michael and He, Yuxiong},
  journal={arXiv preprint arXiv:2507.11830},
  year={2025}
}

@misc{shift_parallelism,
  title        = {Shift Parallelism: Low-Latency, High-Throughput LLM Inference for Dynamic Workloads},
  author       = {Hidayetoglu, Mert and Qiao, Aurick and Wyatt, Michael and Rasley, Jeff and He, Yuxiong and Rajbhandari, Samyam},
  year         = {2025},
  month        = sep,
  eprint       = {2509.16495},
  archivePrefix= {arXiv},
  primaryClass = {cs.DC},
  doi          = {10.48550/arXiv.2509.16495},
  url          = {https://arxiv.org/abs/2509.16495}
}

@article{xu2025ultralong,
  title   = {From 128K to 4M: Efficient Training of Ultra-Long Context Large Language Models},
  author  = {Xu, Chejian and Ping, Wei and Xu, Peng and Liu, Zihan and Wang, Boxin and Shoeybi, Mohammad and Li, Bo and Catanzaro, Bryan},
  journal = {arXiv preprint arXiv:2504.06214},
  year    = {2025}
}

@article{llama3_herd,
  title   = {The Llama 3 Herd of Models},
  author  = {Grattafiori, Aaron and others},
  journal = {arXiv preprint arXiv:2407.21783},
  year    = {2024},
  url     = {https://arxiv.org/abs/2407.21783}
}

@article{gptoss_modelcard,
  title   = {gpt-oss-120b \& gpt-oss-20b Model Card},
  author  = {Agarwal, Shyamal and others},
  journal = {arXiv preprint arXiv:2508.10925},
  year    = {2025},
  url     = {https://arxiv.org/abs/2508.10925}
}

@inproceedings{agrawal2024taming,
  title={Taming $\{$Throughput-Latency$\}$ tradeoff in $\{$LLM$\}$ inference with $\{$Sarathi-Serve$\}$},
  author={Agrawal, Amey and Kedia, Nitin and Panwar, Ashish and Mohan, Jayashree and Kwatra, Nipun and Gulavani, Bhargav and Tumanov, Alexey and Ramjee, Ramachandran},
  booktitle={18th USENIX Symposium on Operating Systems Design and Implementation (OSDI 24)},
  pages={117--134},
  year={2024}
}

@misc{vllm_metrics,
  title        = {Metrics},
  author       = {{vLLM Project}},
  howpublished = {\url{https://docs.vllm.ai/en/stable/design/metrics/}}
}

@misc{prometheus_io,
  title = {Prometheus - Monitoring system \& time series database},
  author = {{The Prometheus Authors}},
  year = {2025},
  url = {https://prometheus.io/},
  howpublished = {\url{https://prometheus.io/}},
}

@article{elradi2025prometheus,
  title={Prometheus \& Grafana: A Metrics-focused Monitoring Stack},
  author={Elradi, Mohammed Daffalla},
  journal={Journal of Computer Allied Intelligence (JCAI, ISSN: 2584-2676)},
  volume={3},
  number={3},
  pages={28--39},
  year={2025}
}

@article{ainslie2023gqa,
  title={Gqa: Training generalized multi-query transformer models from multi-head checkpoints},
  author={Ainslie, Joshua and Lee-Thorp, James and De Jong, Michiel and Zemlyanskiy, Yury and Lebr{\'o}n, Federico and Sanghai, Sumit},
  journal={arXiv preprint arXiv:2305.13245},
  year={2023}
}

@article{agrawal2023sarathi,
  title={Sarathi: Efficient llm inference by piggybacking decodes with chunked prefills},
  author={Agrawal, Amey and Panwar, Ashish and Mohan, Jayashree and Kwatra, Nipun and Gulavani, Bhargav S and Ramjee, Ramachandran},
  journal={arXiv preprint arXiv:2308.16369},
  year={2023}
}

@article{li2022competition,
  title={Competition-level code generation with alphacode},
  author={Li, Yujia and Choi, David and Chung, Junyoung and Kushman, Nate and Schrittwieser, Julian and Leblond, R{\'e}mi and Eccles, Tom and Keeling, James and Gimeno, Felix and Dal Lago, Agustin and others},
  journal={Science},
  volume={378},
  number={6624},
  pages={1092--1097},
  year={2022},
  publisher={American Association for the Advancement of Science}
}

@article{team2025kimi,
  title={Kimi k2: Open agentic intelligence},
  author={Team, Kimi and Bai, Yifan and Bao, Yiping and Chen, Guanduo and Chen, Jiahao and Chen, Ningxin and Chen, Ruijue and Chen, Yanru and Chen, Yuankun and Chen, Yutian and others},
  journal={arXiv preprint arXiv:2507.20534},
  year={2025}
}

@article{team2023gemini,
  title={Gemini: a family of highly capable multimodal models},
  author={Team, Gemini and Anil, Rohan and Borgeaud, Sebastian and Alayrac, Jean-Baptiste and Yu, Jiahui and Soricut, Radu and Schalkwyk, Johan and Dai, Andrew M and Hauth, Anja and Millican, Katie and others},
  journal={arXiv preprint arXiv:2312.11805},
  year={2023}
}

@misc{wang2024executable,
      title={Executable Code Actions Elicit Better LLM Agents}, 
      author={Xingyao Wang and Yangyi Chen and Lifan Yuan and Yizhe Zhang and Yunzhu Li and Hao Peng and Heng Ji},
      year={2024},
      eprint={2402.01030},
      archivePrefix={arXiv},
      primaryClass={cs.CL}
}

@article{chen2021evaluating,
  title={Evaluating large language models trained on code},
  author={Chen, Mark},
  journal={arXiv preprint arXiv:2107.03374},
  year={2021}
}

@article{hu2023chatgpt,
  title={ChatGPT sets record for fastest-growing user base-analyst note},
  author={Hu, Krystal and others},
  journal={Reuters},
  volume={12},
  pages={2023},
  year={2023}
}

@misc{nccl,
  title        = {{NCCL}: NVIDIA Collective Communications Library},
  author       = {{NVIDIA Corporation}},
  year         = {2017},
  howpublished = {\url{https://github.com/NVIDIA/nccl}},
  note         = {}
}

@misc{gloo,
  title        = {Gloo: Collective Communications Library},
  author       = {{PyTorch Contributors}},
  year         = {2017},
  howpublished = {\url{https://github.com/pytorch/gloo}},
  note         = {}
}

@article{mo2025hetis,
  title={Hetis: Serving LLMs in Heterogeneous GPU Clusters with Fine-grained and Dynamic Parallelism},
  author={Mo, Zizhao and Liao, Jianxiong and Xu, Huanle and Zhou, Zhi and Xu, Chengzhong},
  journal={arXiv preprint arXiv:2509.08309},
  year={2025}
}

@article{seesaw2025,
  title={Seesaw: High-throughput LLM Inference via Model Re-sharding},
  author={Su, Qidong and Zhao, Wei and Li, Xin and Andoorveedu, Muralidhar and Jiang, Chenhao and Zhu, Zhanda and Song, Kevin and Giannoula, Christina and Pekhimenko, Gennady},
  journal={arXiv preprint arXiv:2503.06433},
  year={2025}
}

@article{liu2025hap,
  title={HAP: Hybrid Adaptive Parallelism for Efficient Mixture-of-Experts Inference},
  author={Liu, Aixin and others},
  journal={arXiv preprint arXiv:2508.19373},
  year={2025}
}

@article{chen2025gyges,
  title={Gyges: Dynamic Cross-Instance Parallelism Transformation for Efficient LLM Inference},
  author={Chen, Haoyu and Li, Xue and Qian, Kun and Wang, Xin},
  journal={arXiv preprint arXiv:2509.19729},
  year={2025}
}

@inproceedings{wagenlander2024tenplex,
  title={Tenplex: Dynamic Parallelism for Deep Learning using Parallelizable Tensor Collections},
  author={Wagenl{\"a}nder, Marcel and Li, Guo and Zhao, Bo and Mai, Luo and Pietzuch, Peter R},
  booktitle={Proceedings of the 30th Symposium on Operating Systems Principles (SOSP '24)},
  year={2024}
}

@article{wu2023fastserve,
  title={Fast Distributed Inference Serving for Large Language Models},
  author={Wu, Bingyang and others},
  journal={arXiv preprint arXiv:2305.05920},
  year={2023}
}

@inproceedings{li2023alpaserve,
  title={AlpaServe: Statistical Multiplexing with Model Parallelism for Deep Learning Serving},
  author={Li, Zhuohan and others},
  booktitle={17th USENIX Symposium on Operating Systems Design and Implementation (OSDI 23)},
  year={2023}
}

@inproceedings{wu2024loongserve,
  title={Loongserve: Efficiently serving long-context large language models with elastic sequence parallelism},
  author={Wu, Bingyang and Liu, Shengyu and Zhong, Yinmin and Sun, Peng and Liu, Xuanzhe and Jin, Xin},
  booktitle={Proceedings of the ACM SIGOPS 30th Symposium on Operating Systems Principles},
  pages={640--654},
  year={2024}
}

@inproceedings{brown2020language,
  title={Language models are few-shot learners},
  author={Brown, Tom and Mann, Benjamin and Ryder, Nick and Subbiah, Melanie and Kaplan, Jared D and Dhariwal, Prafulla and Neelakantan, Arvind and Shyam, Pranav and Sastry, Girish and Askell, Amanda and others},
  booktitle={Advances in Neural Information Processing Systems},
  volume={33},
  pages={1877--1901},
  year={2020}
}

@article{touvron2023llama,
  title={Llama: Open and efficient foundation language models},
  author={Touvron, Hugo and Lavril, Thibaut and Izacard, Gautier and Martinet, Xavier and Lachaux, Marie-Anne and Lacroix, Timoth{\'e}e and Rozi{\`e}re, Baptiste and Goyal, Naman and Hambro, Eric and Azhar, Faisal and others},
  journal={arXiv preprint arXiv:2302.13971},
  year={2023}
}

@article{pope2023efficiently,
  title={Efficiently scaling transformer inference},
  author={Pope, Reiner and Douglas, Sholto and Chowdhery, Aakanksha and Devlin, Jacob and Bradbury, James and Heek, Jonathan and Xiao, Kefan and Agrawal, Shivani and Dean, Jeff},
  journal={Proceedings of machine learning and systems},
  volume={5},
  pages={606--624},
  year={2023}
}

@article{shoeybi2019megatron,
  title={Megatron-lm: Training multi-billion parameter language models using model parallelism},
  author={Shoeybi, Mohammad and Patwary, Mostofa and Puri, Raul and LeGresley, Patrick and Casper, Jared and Catanzaro, Bryan},
  journal={arXiv preprint arXiv:1909.08053},
  year={2019}
}

@inproceedings{kwon2023efficient,
  title={Efficient memory management for large language model serving with pagedattention},
  author={Kwon, Woosuk and Li, Zhuohan and Zhuang, Siyuan and Sheng, Ying and Zheng, Lianmin and Yu, Cody Hao and Gonzalez, Joseph and Zhang, Hao and Stoica, Ion},
  booktitle={Proceedings of the 29th symposium on operating systems principles},
  pages={611--626},
  year={2023}
}

@article{hidayetoglu2025shift,
  title={Shift Parallelism: Low-Latency, High-Throughput LLM Inference for Dynamic Workloads},
  author={Hidayetoglu, Mert and Qiao, Aurick and Wyatt, Michael and Rasley, Jeff and He, Yuxiong and Rajbhandari, Samyam},
  journal={arXiv preprint arXiv:2509.16495},
  year={2025}
}

\end{document}